\documentclass[twocolumn]{aastex63}

\shorttitle{Simulated UFDs Around the Milky Way}
\shortauthors{Applebaum et al.}
\graphicspath{{./}{figures/}}

\usepackage[hang]{footmisc}
\begin{document}

\title{Ultra-faint dwarfs in a Milky Way context:\\ Introducing the Mint Condition DC Justice League Simulations}

\correspondingauthor{Elaad Applebaum}
\email{applebaum@physics.rutgers.edu}

\author[0000-0001-8301-6152]{Elaad Applebaum}
\affiliation{Department of Physics and Astronomy, 
Rutgers, The State University of New Jersey, 
136 Frelinghuysen Rd., 
Piscataway, NJ 08854, USA}

\author[0000-0002-0372-3736]{Alyson M. Brooks}
\affiliation{Department of Physics and Astronomy, 
Rutgers, The State University of New Jersey, 
136 Frelinghuysen Rd., 
Piscataway, NJ 08854, USA}

\author[0000-0001-6779-3429]{Charlotte R. Christensen}
\affiliation{Physics Department, Grinnell College,
1116 Eighth Avenue, Grinnell, IA 50112, USA}

\author[0000-0002-9581-0297]{Ferah Munshi}
\affiliation{Department of Physics \& Astronomy, University of Oklahoma, 440 W. Brooks St., Norman, OK 73019, USA}

\author{Thomas R. Quinn}
\affiliation{Department of Astronomy, University of Washington, Box 351580, Seattle, WA, 98115, USA}

\author[0000-0001-8523-1171]{Sijing Shen}
\affiliation{Institute of Theoretical Astrophysics, University of Oslo, Postboks 1029, 0315 Oslo, Norway}

\author[0000-0002-4353-0306]{Michael Tremmel}
\affiliation{Physics Department, Yale Center for Astronomy \& Astrophysics, 
PO Box 208120, New Haven, CT 06520, USA}



\begin{abstract}

We present results from the ``Mint'' resolution DC Justice League suite of Milky Way-like zoom-in cosmological simulations, which extend our study of nearby galaxies down into the ultra-faint dwarf (UFD) regime for the first time. The mass resolution of these simulations is the highest ever published for cosmological Milky Way zoom-in simulations run to $z=0$, with initial star (dark matter) particle masses of 994~(17900)~M$_\sun$, and a force resolution of 87~pc. We study the surrounding dwarfs and UFDs, and find the simulations match the observed dynamical properties of galaxies with $-3 < M_V < -19$, and reproduce the scatter seen in the size-luminosity plane for $r_h\gtrsim200$~pc. We predict the vast majority of nearby galaxies will be observable by the Vera Rubin Observatory's co-added Legacy Survey of Space and Time (LSST). We additionally show that faint dwarfs with velocity dispersions $\lesssim5$~km/s result from severe tidal stripping of the host halo. We investigate the quenching of UFDs in a hydrodynamical Milky Way context, and find that the majority of UFDs are quenched prior to interactions with the Milky Way, though some of the quenched UFDs retain their gas until infall. Additionally these simulations yield some unique dwarfs that are the first of their kind to be simulated, e.g., an HI-rich field UFD, a late-forming UFD that has structural properties similar to Crater 2, as well as a compact dwarf satellite that has no dark matter at $z=0$.

\end{abstract}



\section{Introduction}\label{sec:intro}

In recent years, many simulations have focused on the dwarf galaxy regime to test our understanding of galaxy formation. Not only are dwarf galaxies the closest galaxies to the Milky Way, but their smaller potential wells make them more sensitive tests of our physical models.

Most dwarf galaxy simulations have focused on galaxies with M$_\mathrm{star}\gtrsim10^{5-6}$~M$_\sun$, in the mass range of the Milky Way's ``classical dwarf'' satellite galaxies. With these simulations, we have greatly improved our understanding of galaxy formation, thanks to advances in resolution, the detailed modeling of relevant physical processes, and a consideration of observational biases. For example, simulations in a $\Lambda$CDM universe can now explain the number, distribution, and central densities of classical Milky Way satellites \citep[e.g.,][]{Zolotov2012, Brooks2013, Brooks2014, Wetzel2016, Sawala2016, Tomozeiu2016, Santos-Santos2018, Garrison-Kimmel2019}. Various simulations explain both the diversity of dwarf galaxy star formation histories in the Local Group as well as average mass-dependent trends \cite[e.g.,][]{Benitez-Llambay2015, Wetzel2016, Wright2019, Buck2019, Digby2019, Garrison-Kimmel2019b}. Additionally, many simulations reproduce a variety of other scaling relations in this mass range, such as the stellar mass-halo mass, Tully-Fisher, and mass-metallicity relations \cite[e.g.,][]{Munshi2013, Vogelsberger2013, Shen2014, Christensen2016, Christensen2018, Brook2016, Brooks2017,  El-Badry2018, Santos-Santos2018}. While work is ongoing and many questions remain unsettled, such as the radial \citep[e.g.,][]{Samuel2020} or planar \citep[e.g.,][]{Ahmed2017} distributions of classical dwarf satellites, our ability to model galaxies in the classical dwarf regime has dramatically improved in the last decade, and we have successfully explained a variety of properties of observed galaxies.

The advent of digital sky surveys has led to the rapid discovery of dozens of new dwarf galaxies around the Milky Way \citep[see][for a recent review]{Simon2019}, largely in the regime of the ultra-faint dwarfs (UFDs; $M_V$ fainter than $-8$\footnote{We use $M_V=-8$ as our boundary, but note that \citet{Simon2019} has argued for $M_V=-7.7$.}, and M$_\mathrm{star}\lesssim10^5$~M$_\sun$). In this sub-classical regime, however, our understanding is incomplete, and more work must be done to replicate the successes seen in simulating higher mass dwarfs.

Given the pace of discovery, there is still a large uncertainty in the number and distribution of these faint dwarfs. Different assumptions about survey completeness and the underlying halo distribution lead to estimates differing by nearly an order of magnitude in the predicted number of satellites \citep{Simon2007, Tollerud2008, Hargis2014, Newton2018, Jethwa2018, Drlica-Wagner2020, Nadler2020}. Predictions for the Milky Way satellite distribution are influenced by uncertainties in the connection between halos and galaxies, such as the relationship between stellar mass and halo mass in small halos \citep[e.g.,][]{Garrison-Kimmel2017, Munshi2017, Read2019, Rey2019}, or the surface brightnesses---and therefore detectability---of galaxies in low-mass halos \citep[e.g.][]{Bullock2010, Wheeler2019}. Differing assumptions about which halos can host galaxies can even lead to a ``too few satellites'' problem \citep{Kim2018, Graus2019}, in which there are more Milky Way satellites than theoretically expected, reversing the decades-old Missing Satellites Problem \citep{Klypin1999, Moore1999}.

The star formation histories (SFHs) and quenching mechanisms of UFD galaxies are also uncertain. It has been suggested that ultra-faint dwarf galaxies are fossils of reionization \citep{Bovill2009}, having been quenched via gas heating during reionization \citep[e.g.,][]{Bullock2000, Benson2002, Somerville2002}. Early quenching is consistent with observations of some UFDs \citep[e.g.,][]{Brown2014, Weisz2014a}, but all UFDs with constrained star formation histories are close to the Milky Way or M31, complicating any interpretation. Previous simulations of isolated UFDs \citep[e.g.,][]{Fitts2017, Jeon2017, Wheeler2019} are consistent with reionization quenching. However, simulations must be able to simultaneously explain the apparent early quenching of most UFDs, along with the existence of UFDs hosting recent star formation, such as Leo T (\citealt{Irwin2007}; see also \citealt{Rey2020}).

Other properties of the newly discovered nearby faint dwarfs are becoming clearer, including their kinematics \citep[e.g.,][]{Kleyna2005, Munoz2006, Simon2007, Wolf2010, Koposov2011, Kirby2013a}, morphology and structure \citep[e.g.,][]{Martin2008, McConnachie2012, Munoz2018}, and metallicity and chemical composition \citep[e.g.,][]{Simon2007, Frebel2010, Norris2010, Vargas2013, Kirby2013b, Ji2020}. As our knowledge of faint galaxies increases, the emerging view is that below the mass of classical dwarfs, galaxies trend towards increasingly ancient and dark matter-dominated stellar systems. Even UFD galaxies seem to be in many ways a natural extension of more luminous systems to lower mass, with any clear physical division likely to be driven by the details of reionization \citep{Bose2018, Simon2019}. Nonetheless, even among the faintest dwarfs, there is a great deal of galaxy-to-galaxy diversity, including in kinematics, sizes, and star formation histories, that has proven challenging to reproduce in existing simulations.

Now that dozens of new galaxies have been discovered around the Milky Way, it is crucial to test our galaxy formation models in this fainter regime, and to ensure that we can still match and explain the properties of observed dwarf galaxies. However, while there are a wealth of Milky Way simulations resolving classical dwarf galaxies, there is a paucity of simulations capable of resolving down to the UFD range.

It is important, therefore, to run new simulations capable of resolving the Milky Way's fainter satellites. However, it is computationally expensive to achieve the resolution necessary to resolve down to the UFD range while simultaneously placing galaxies in a cosmological context allowing for gas inflow and outflow as well as tidal interactions with larger galaxies. As alternatives, several groups have undertaken direct simulation of very small dwarf galaxies in non-cosmological contexts \citep[e.g.,][]{Read2016, Corlies2018, Emerick2019}. Other groups have simulated cosmological regions at high resolution, but have stopped at high redshift \citep[e.g.,][]{Wise2014, Jeon2015, Safarzadeh2017, Maccio2017}, or used the results as initial conditions for later host-satellite simulations \citep{Frings2017}. Finally, there have been several simulations of field dwarfs in cosmological environments, achieving analogs to dwarf galaxies far from the Milky Way \citep[e.g.,][]{Simpson2013, Munshi2017, Munshi2019, Onorbe2015, Wheeler2015, Wheeler2019, Jeon2017,Fitts2017, Revaz2018, Agertz2020}.  
Cosmological simulations have made significant strides in resolution \citep[e.g.,][]{VINT1, VINT2, VINT3}, but have not achieved the mass resolution required to reliably study the properties of galaxies with M$_\mathrm{star}\lesssim10^5$~M$_\sun$ in a Milky Way context \citep[e.g.,][]{Zolotov2012, Brooks2014, Sawala2016, Wetzel2016, Simpson2018, Buck2019, Garrison-Kimmel2019}.
To test whether our models still match observations given our burgeoning Milky Way census, it will be necessary to achieve higher resolution in a Milky Way context.

To this end, we introduce the DC Justice League suite of Milky Way zoom-in simulations, run at high (``Mint'') resolution sufficient to begin probing analogs of the faintest Milky Way satellites. While our studies of spatially resolved galaxies are limited to larger UFDs, these simulations serve as a crucial step forward in our study of the Milky Way environment. We will describe the global properties of galaxies as faint as $M_V\sim-3$ and the resolved properties of dwarfs with $M_V\lesssim-5$. We present two simulations run from $z=159$ to $z=0$, with present-day Milky Way halo masses of $7.5\times10^{11}$~M$_\sun$ and $2.4\times10^{12}$~M$_\sun$, allowing us to bracket the suspected lower and upper limits of the Milky Way's mass, respectively. We use these simulations to show that we can match dwarf galaxy properties simultaneously across 6 orders of magnitude in luminosity, including lower luminosities than ever before studied around a fully cosmological Milky Way simulation. We further take advantage of these new simulations to study the star formation histories and gas properties of UFDs around the Milky Way. We focus in particular on the question of what quenched star formation in UFDs, which in previous simulations could not be studied in the context of the Milky Way. Through case studies, we finally show how much of the variety seen in faint galaxy properties arises naturally in our simulations.

The paper is organized as follows: in Section~\ref{sec:sims} we describe our simulations. In Section~\ref{sec:centrals} we present the basic properties of the Milky Way-like galaxies. We then discuss the properties of the dwarf galaxies in Section~\ref{sec:dwarfs}. In Section~\ref{sec:quenching} we show that reionization is responsible for quenching the majority of UFD galaxies, even around the Milky Way. We present several case studies of interesting galaxies in Section~\ref{sec:casestudies}. We discuss our results in Section~\ref{sec:discussion}, including limitations of this work. We summarize our results in Section~\ref{sec:summary}.

\section{Simulations}\label{sec:sims}

The simulations used in this work were run using \textsc{ChaNGa} \citep{Menon2015}, a smoothed particle hydrodynamics (SPH) + N-body code. \textsc{ChaNGa} includes the hydrodynamic modules of \textsc{Gasoline} \citep{Wadsley2004, Wadsley2017} but uses the \textsc{charm++} \citep{Kale1993} runtime system for dynamic load balancing and communication to allow scalability up to thousands of cores. \textsc{ChaNGa} also incorporates an improved gravity solver that is intrinsically faster than \textsc{Gasoline}.

The simulations were run using the ``zoom-in'' technique \citep[e.g.,][]{Katz1993, Onorbe2014}, where smaller regions within large, dark matter-only volumes are resimulated at higher resolution with full hydrodynamics. The zoom-in technique allows for very high resolutions in the regions of interest, while still capturing large-scale gravitational tidal torques. The zoom regions were selected from a 50~Mpc, dark matter-only volume run using the \citet{Planck2016} cosmological parameters. The high-resolution regions are largely uncontaminated by low-resolution particles out to 2~R$_\mathrm{vir}$ for each host. However, since the zoom regions are non-spherical we find galaxies out to ${\sim}$2.5~R$_\mathrm{vir}$ in the present day. Gas particles are split from the dark matter particles according to the cosmic baryon fraction $\Omega_\mathrm{bar}/\Omega_\mathrm{m}=0.156$. The present-day central halos were chosen to be Milky Way analogs; they are isolated and have virial masses bracketing the range of observationally constrained estimates ($\approx0.5-2.5\times10^{12}$~M$_\sun$; e.g., \citealt{Wilkinson1999, Watkins2010, Kafle2014, Sohn2018, Eadie2019}). The simulations have a gravitational spline force softening of 87~pc, minimum hydrodynamical smoothing length of 11~pc, and dark, gas, and (maximum) initial star particle masses of \mbox{17900, 3310, and 994 M$_\sun$}, respectively. These constitute the highest mass resolution of any cosmological simulations ever run of Milky Way-like galaxies. At $z=0$, the two simulations contain approximately $10^{8.3}$ and $10^{8.8}$ particles; in total, they required approximately 14 million and 120 million core hours, respectively. Despite their large computational expense, the simulations were possible owing to the excellent scaling of \textsc{ChaNGa}.

    \begin{deluxetable*}{c c c c c c c c c c}
    \tablecaption{Properties of the simulations.}\label{table:properties}
    \tablewidth{0pt}
        \tablehead{
        \colhead{Simulation} & \colhead{M$_\mathrm{vir}$} & \colhead{R$_\mathrm{vir}$} &
        \colhead{M$_{\mathrm{star},R}$} & \colhead{M$_\mathrm{star,sim}$} 
        & \colhead{$R_d$} &  \colhead{N$_\mathrm{sat}$} & \colhead{N$_\mathrm{field}$} & \colhead{N$_\mathrm{sat,prior}$}\\
         & \colhead{($10^{12}$~M$_\sun$)} & \colhead{(kpc)} & \colhead{($10^{10}$~M$_\sun$)} & \colhead{($10^{10}$~M$_\sun$)} & \colhead{(kpc)} &  
         & & & }
         \startdata
        Elena & 0.75 & 240 & 3.7 & 6.8 & 3.8 & 12 (8) & 5 (2) & 0 (0)\\
        Sandra & 2.4 & 350 & 9.0 & 16.5 & 3.5 & 51 (34) & 18 (16) & 10 (5)\\
        \enddata
        \tablecomments{The name of each simulation, the virial mass (M$_\mathrm{vir}$), virial radius (R$_\mathrm{vir}$), and stellar mass (within $3\times$ the 3D half-mass radius) of the main Milky Way halo, calculated from the $R$-band luminosity assuming a stellar mass-to-light ratio of 1 (M$_{\mathrm{star},R}$) and from the particle data (M$_\mathrm{star,sim}$), the scale length ($R_d$) as found by fitting an exponential profile to the face-on stellar surface mass density in a region beyond the central bulge, the number of satellite galaxies of the main halo (N$_\mathrm{sat}$) that meet both our resolution criteria (globally resolved first, structurally resolved in parentheses; see Section~\ref{sec:sims}), the number of central galaxies beyond the virial radius (N$_\mathrm{field}$) that are globally (structurally) resolved, and the number of present-day Milky Way satellites that fell in as satellites of another dwarf galaxy (N$_\mathrm{sat,prior}$) that are globally (structurally) resolved; N$_\mathrm{sat,prior}$ are a subset of N$_\mathrm{sat}$.}
    \end{deluxetable*}
    
The Milky Way simulation suite presented here serves as a complement to the MARVEL-ous Dwarfs, a suite of four high-resolution zoom-in regions of field dwarf galaxies formed in low-density environments (Munshi~et~al.~in~prep). The Milky Way simulations we discuss here are nicknamed the ``DC Justice League,'' named in honor of the female United States Supreme Court justices. While there are four Milky Way zoom-in simulations in the suite, we discuss two in this paper that have been run at the above-described resolution; we term these ``Mint'' resolution. The two have been nicknamed ``Sandra'' and ``Elena.'' Lower resolution versions of these simulations (run at 175~pc resolution, dubbed ``Near Mint'') have been presented elsewhere \citep{Bellovary2019, Akins2020, Iyer2020}, but we are introducing these high-resolution simulations here for the first time, with spatial and mass resolutions within a factor of ${\sim}2$ of the MARVEL-ous dwarfs (Munshi~et~al.~in~prep).

Star particles represent simple stellar populations with a \citet{Kroupa2001} initial mass function (IMF) and an initial mass of 30\% that of their parent gas particle. We use the ``blastwave'' form of supernova feedback \citep{Stinson2006}, in which mass, metals, and energy from Type II supernovae are deposited among neighboring gas particles. We distribute $1.5\times10^{51}$~erg per supernova, then turn off cooling until the end of the snowplow phase (the extra energy above $10^{51}$~erg is designed to mimic the energy injected into the local ISM by all feedback processes coming from young stars, including high energy radiation). The simulations also incorporate feedback from Type Ia supernovae, mass loss in stellar winds, and iron, oxygen, and total metal enrichment \citep{Stinson2006}, a time-dependent, uniform UV background \citep{Haardt2012}, and metal cooling and diffusion in the interstellar medium \citep{Shen2010}. We discuss the effect of feedback models on our results later in the paper (Section \ref{sec:discussion}).

Star formation in these simulations is based on the local non-equilibrium abundance of molecular hydrogen \citep[H$_2$;][]{Christensen2012}. The recipe follows the creation and destruction of H$_2$ both in the gas-phase and on dust-grains, as well as dissociation via Lyman-Werner radiation. We include both dust-shielding and self-shielding of H$_2$ from radiation, as well as dust-shielding of HI. The probability of forming a star particle of mass $m_\mathrm{star}$ from a gas particle of mass $m_\mathrm{gas}$ is
\begin{equation}
    p=\frac{m_\mathrm{gas}}{m_\mathrm{star}}\left(1-e^{c_0^* X_{\mathrm{H}_2}\Delta t / t_\mathrm{dyn}}\right),
\end{equation}
where $X_{\mathrm{H}_2}$ is the is H$_2$ abundance and $t_\mathrm{dyn}$ is the local dynamical time. The star formation efficiency parameter, $c_0^*=0.1$, is calibrated to provide the correct normalization in the Kennicutt-Schmidt relation \citep{Christensen2014b}. This star formation prescription successfully reproduces the low velocity dispersion of star-forming gas, which is critical to forming the kinematically cold young stars seen in the Milky Way's age-velocity relation \citep{Bird2020}.

We also model supermassive black hole (SMBH) formation, growth, feedback, and dynamics based on local gas conditions \citep{Tremmel2015, Tremmel2017, Bellovary2019}. SMBHs form in cold ($\mathrm{T}<2\times10^4$~K), primordial ($Z<10^{-4}$ and $X_{\mathrm{H}_2}<10^{-4}$), and dense ($n_\mathrm{H}>1.5\times10^4$~cm$^{-3}$) gas, with a seed mass of $5\times10^4$~M$_\sun$. Black holes grow by accreting gas using a modified Bondi-Hoyle formalism that includes a term for momentum supported gas, and by merging with other black holes. Black holes are allowed to move freely within their host galaxies, while explicitly modeling unresolved dynamical friction; this freedom can lead to delayed SMBH mergers \citep{Tremmel2018a, Tremmel2018b} and off-center black holes in dwarf galaxies \citep{Bellovary2019}. Akin to supernova blastwave feedback, SMBHs deposit thermal energy in surrounding gas when they accrete gas, and we turn off cooling in the heated gas for the length of the SMBH time step (usually $<10^4$~yr), with a feedback coupling efficiency of 0.02. We assume accretion is Eddington limited, with a radiative efficiency of 0.1.

We identify halos using \textsc{Amiga's Halo Finder} \citep[AHF;][]{Gill2004, Knebe2009}, which identifies a halo as the spherical region within which the density satisfies a redshift-dependent overdensity criterion based on the approximation of \citet{Bryan1998}. In the case of subhalos embedded in their hosts' density field, their overdensity may never fall below the \citet{Bryan1998} threshold, in which case the extent of the halo is truncated where the radial density profile experiences an upturn. Unbound particles are iteratively removed from all halos, with particles considered unbound if their velocity exceeds the halo escape velocity.

We use AHF for all halo properties unless otherwise stated. Galaxies are defined as all stellar content residing within halos\footnote{For dwarf galaxies, it makes little difference whether all stars or only those within 10\% of the virial radius are considered.}, and satellite galaxies are galaxies residing within subhalos. We trace all main progenitors with at least 100 particles at $z=0$ back in time, and include in our final sample those galaxies with at least 10 star particles and 1000 dark matter particles prior to mass loss due to interactions with the central halo. This corresponds to a peak dark matter halo mass of M$_\mathrm{peak}\ge10^{7.25}$~M$_\sun$. For resolving structural properties of the galaxies, we require at least 50 star particles; therefore, half-light radii and mass-to-light ratios within the half-light radius are only calculated down to the 50 star particle limit, or $M_V\sim-5$, while other properties are calculated across the entire sample, down to $M_V\sim-3$. Table~\ref{table:properties} shows the number of galaxies in our sample that meet our resolution criteria, along with the basic properties of the two Milky Way-like halos.

Further analysis was performed using the \textsc{pynbody} analysis code \citep{pynbody}. Galaxy magnitudes and luminosities are calculated by interpolating on a grid of metallicities and ages, using Padova simple stellar population isochrones \citep{Marigo2008, Girardi2010}\footnote{\url{http://stev.oapd.inaf.it/cgi-bin/cmd}}. We make no corrections for dust extinction; we expect dust to have little impact in the dwarf galaxy regime focused on in this work.

Simulations run with \textsc{ChaNGa} and \textsc{Gasoline} using the above star formation and feedback models have yielded numerous results in the dwarf galaxy regime, and have explained a variety of observed properties, such as the stellar mass-halo mass relation \citep{Munshi2013, Munshi2017}, the baryonic Tully-Fisher relation \citep{Christensen2016, Brooks2017}, the mass-metallicity relation \citep{Brooks2007, Christensen2018}, and the properties of Milky Way satellites \citep{Zolotov2012, Brooks2014} and field dwarfs \citep{Brooks2017}. These models produced the first simulated cored dark matter density profiles and bulgeless disk galaxies \citep{Governato2010, Brook2011, Governato2012}. The simulations have also been used to make observable predictions for the star formation histories of nearby dwarf galaxies \citep{Wright2019} and the merger rates of dwarf galaxy SMBHs \citep{Bellovary2019}.

\section{Milky Way-like Galaxies}\label{sec:centrals}

While the focus of this work is on the satellite and other nearby dwarf galaxies, we also briefly present the properties of the central, Milky Way-like galaxies. Future work will investigate these galaxies more closely.

\begin{figure}
    \epsscale{1.1}
    \plotone{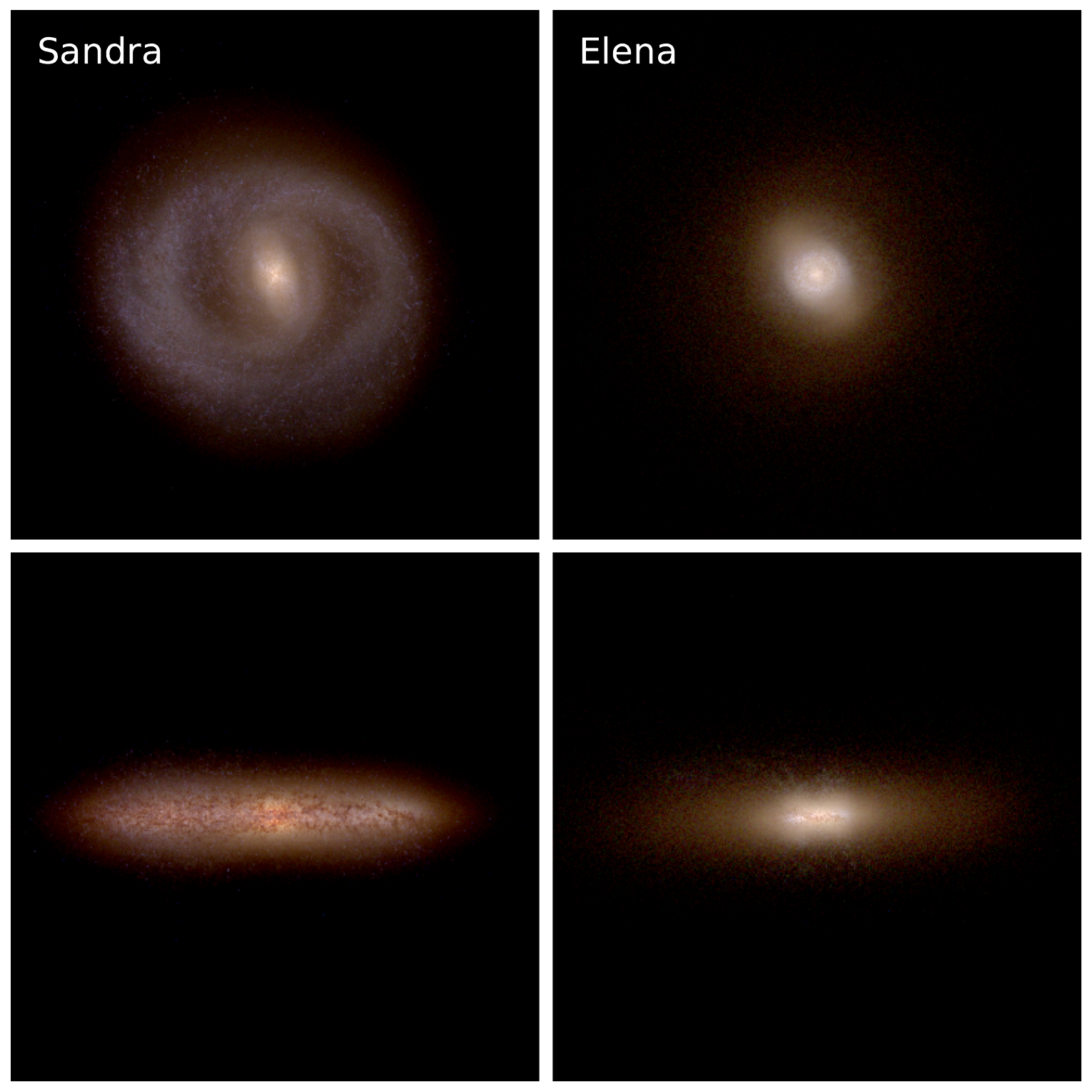}
    \caption{Mock UVI images of Sandra (left column) and Elena (right column), for both face-on (top) and edge-on (bottom) orientations. Images were generated using outputs from the Monte Carlo radiative transfer code SKIRT, assuming a dust-to-metals ratio of 0.3 and a maximum dust temperature of 8000~K. Images are 40~kpc across. Elena is shown to a dimmer surface brightness (23~mag~arcsec${}^{-2}$) than Sandra (21~mag~arcsec${}^{-2}$) in order to highlight the low surface brightness disk. Sandra shows a strong central bar, flocculent spiral arms, and a dusty disk. Elena shows an apparent ring structure and an extended low surface brightness disk.}
    \label{fig:centrals}
\end{figure}

Figure~\ref{fig:centrals} shows mock face-on and edge-on multi-band images of the galaxies. These images have been generated using the Monte Carlo radiative transfer code SKIRT \citep{Baes2003, Baes2011, Camps2020}, assuming a dust-to-metals ratio of 0.3. The two galaxies have very different morphology. Sandra has flocculent spiral arms and a clear bar structure in the center. Elena, on the other hand, hosts no spiral arms but has a star-forming ring. As late as $z\sim0.5$, Elena had spiral and bar structures. However, during its latest merger (see below), it began to quench and redden, and its morphology transformed to the one seen in the figure. While it may not be morphologically a Milky Way analogue, its halo and stellar masses, as well as its relatively quiet assembly history, are thought to be consistent with that of the Milky Way. We will therefore continue to refer to it as a Milky Way-like galaxy.

Summary properties of the two Milky Way-like galaxies, including virial masses, virial radii, and number of dwarf satellites, are listed in Table~\ref{table:properties}. We include the stellar masses within $3\times$ the 3D half-mass radii, as derived from the $R$-band luminosity assuming a stellar mass-to-light ratio of 1, and directly from the particle data. The masses are higher than expected from abundance matching \citep[e.g.][]{Moster2013}, but the photometrically derived masses are ${\sim}40$\% lower, similar to the results of \citet{Munshi2013}. Thus, it is unclear if the Milky Way galaxies are suffering from overcooling. If so, it is not a result of the increased resolution, as the Mint resolution stellar masses are within 10\% of the Near Mint runs. We will explore this in the future using \textsc{ChaNGa}'s superbubble feedback recipe \citep{Keller2014}, which has been shown to drive more efficient outflows.

Throughout its history, Sandra experiences multiple mergers with LMC-mass halos\footnote{For the sake of brevity, we are considering halos with virial mass between $8\times10^{10}$~M$_\sun$ and $2\times10^{11}$~M$_\sun$ at infall to be ``LMC-mass.''}. Its last major merger is with an LMC-mass halo (merger ratio ${\sim}1.5$) at $z\sim2$, though the first infall of the galaxy occurs earlier, at $z\sim3$. During this time in the galaxy's history, a clear disk has not yet formed, and many simultaneous mergers occur close in time. Therefore, there is some uncertainty on the exact timing and masses involved. However, the merger is consistent with a \textit{Gaia-Enceladus}/Sausage-like event \citep[e.g.,][]{Belokurov2018, Helmi2018}. By $z\sim1.5$, a clear disk forms, around which time another LMC-mass halo falls in (merger ratio $\approx7$), completing its merger by $z=1$. At $z\sim0.5$, the galaxy experiences another LMC-mass infall (merger ratio $\approx8$), which orbits for several Gyr before merging at $z=0.15$. Finally, in the present-day, an LMC-mass halo satellite is completing its first pericentric passage, currently at a galactocentric distance of 200~kpc.

\begin{figure}
    \epsscale{1.1}
    \plotone{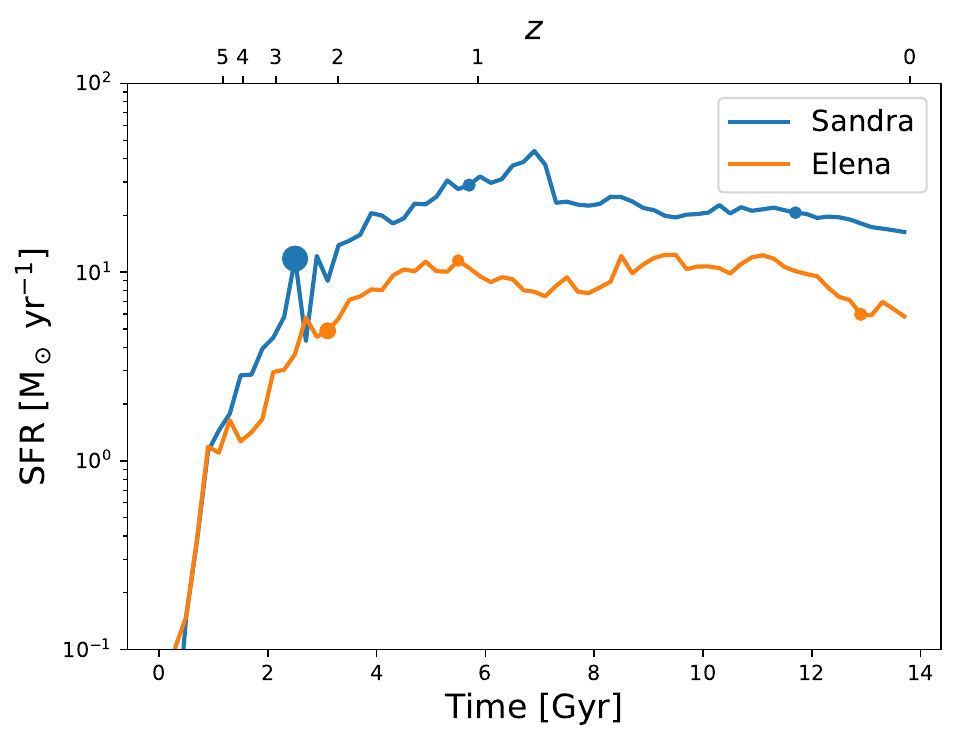}
    \caption{Star formation histories of the 2 Milky Way-like galaxies in the suite, in bins of 200~Myr, calculated for all stars within 3 times their 3D half-mass radii. For each galaxy, the three most major mergers are marked, with the size inversely proportional to the merger ratio (i.e. the largest point is the most major merger). Sandra has a higher star formation rate across most of cosmic time, commensurate with its higher mass. In the last ${\sim}3$~Gyr, Elena has had a declining star formation rate, leading to its redder color. During this time, the central galaxy is harassed by the orbiting dwarf that ultimately merges at ${\sim}13$~Gyr.}
    \label{fig:mwsfh}
\end{figure}

Commensurate with its lower mass, Elena experiences mergers with smaller halos than Sandra. At $z\sim3$, it experiences its most major merger (though, similar to Sandra, there are many simultaneous mergers that complicate the picture), with a merger ratio of 4. At $z\sim1$ it experiences a Sequoia-like infall \citep[e.g.,][]{Barba2019, Myeong2019} of a ${\sim}10^{10}$~M$_\sun$ halo (merger ratio $\approx8$). Finally, at $z\sim0.5$ two unassociated halos fall in, one LMC-mass (merger ratio $\approx7$) and one SMC-mass, with both eventually merging by $z=0$. During their several Gyr of orbit, the two galaxies fly by the Milky Way numerous times and harass it substantially, ultimately leading to the quenching and morphological transition mentioned above. Additionally, about 2~Gyr before the present day, one of these galaxies passes directly through the center of the main galaxy, which may explain its ring structure \citep[see, e.g.,][]{Appleton1996}.

Figure~\ref{fig:mwsfh} shows the star formation histories of the two central galaxies. As expected, Sandra, the more massive galaxy, has a higher star formation rate throughout its history. As noted above, Elena's last merger caused a decline in star formation, visible in the last 2-3~Gyr.

\section{The Dwarf Galaxy Population}\label{sec:dwarfs}

In this section we focus on the properties of the dwarf galaxies in the simulations. First, we discuss the general attributes of the population, demonstrating consistency with observations across the entire luminosity range. Properties of the galaxies that are presented below are collated in Table~\ref{table:galaxies}, the full version of which is available as supplementary material.

\subsection{Observational Sample}\label{sec:obssample}

\begin{deluxetable*}{c c c c c c c c c c c c}[hb]
\tablecaption{Properties of individual dwarf galaxies.}\label{table:galaxies}
\tablewidth{0pt}
\tablehead{
\colhead{Simulation} & \colhead{Halo Number} & \colhead{M$_\mathrm{vir}$} & \colhead{M$_\mathrm{peak}$} & \colhead{M$_\mathrm{star}$} & \colhead{M$_\mathrm{gas}$} & \colhead{M$_\mathrm{HI}$}
 & \colhead{$M_V$} & \colhead{$r_h$} & \colhead{[Fe/H]} & \colhead{R$_\mathrm{gal}$} & \colhead{$\tau_{90}$}\\
\colhead{} & \colhead{} & \colhead{(M$_\sun$)} & \colhead{(M$_\sun$)} & \colhead{(M$_\sun$)} & \colhead{(M$_\sun$)} & \colhead{(M$_\sun$)} & \colhead{} & \colhead{(pc)} & \colhead{} & \colhead{(kpc)} & \colhead{(Gyr)}}
\startdata 
Sandra & 2 & $9.1\times10^{10}$ & $1.3\times10^{11}$ & $2.0\times10^{9}$ & $6.5\times10^{9}$ & $2.3\times10^{9}$ & -18.5 & 2730 & -0.9 & 203 & 13.0${}^{*}$ \\
Sandra & 3 & $4.4\times10^{10}$ & $7.0\times10^{10}$ & $1.4\times10^{9}$ & $1.6\times10^{9}$ & $6.3\times10^{8}$ & -17.4 & 2310 & -0.9 & 235 & 11.5${}^{*}$ \\
Sandra & 4 & $3.1\times10^{10}$ & $3.9\times10^{10}$ & $3.2\times10^{8}$ & $9.1\times10^{8}$ & $3.1\times10^{8}$ & -15.8 & 2370 & -1.2 & 255 & 11.1${}^{*}$ \\
Sandra & 5 & $3.0\times10^{10}$ & $4.2\times10^{10}$ & $2.3\times10^{8}$ & $9.5\times10^{8}$ & $3.5\times10^{8}$ & -15.5 & 1530 & -1.3 & 337 & 11.3${}^{*}$ \\
Elena & 9 & $1.5\times10^{9}$ & $7.6\times10^{9}$ & $2.6\times10^{7}$ & $1.2\times10^{8}$ & $3.1\times10^{7}$ & -13.3 & 2370 & -1.6 & 71 & 11.7${}^{*}$ \\
Elena & 14 & $1.9\times10^{9}$ & $3.3\times10^{9}$ & $2.1\times10^{6}$ & $2.0\times10^{7}$ & $1.1\times10^{6}$ & -10.0 & 665 & -2.0 & 524 & 8.5${}^{*}$ \\
Elena & 16 & $1.7\times10^{9}$ & $3.9\times10^{9}$ & $2.5\times10^{5}$ & $8.6\times10^{5}$ & $4.0\times10^{4}$ & -8.1 & 227 & -2.1 & 200 & 11.5 \\
Elena & 20 & $1.1\times10^{9}$ & $2.0\times10^{9}$ & $6.0\times10^{5}$ & 0.0 & 0.0 & -8.4 & 313 & -2.2 & 157 & 2.3 \\
\enddata

\tablecomments{The Milky Way simulation in which the galaxies are found, their halo numbers in the simulation, their virial masses (M$_\mathrm{vir}$), peak halo masses (M$_\mathrm{peak}$), stellar masses (M$_\mathrm{star}$), total gas masses (M$_\mathrm{gas}$), and HI masses (M$_\mathrm{HI}$), their $V$-band magnitudes and half-light radii ($r_h$; see Section~\ref{sec:structural}), their metallicities, their galactocentric distances (R$_\mathrm{gal}$), and the time at which 90\% of their stars had formed ($\tau_{90}$; if they are still star-forming, $\tau_{90}$ is marked with an asterisk). This table is published in its entirety in machine-readable format, with the four most massive dwarf galaxies from each simulation shown here as examples.}

\end{deluxetable*}

We compare the results of our simulations to several dwarf galaxy catalogs that have been assembled in the literature. In particular, we compare to the updated version of the \citet{McConnachie2012} catalog\footnote{\url{http://www.astro.uvic.ca/~alan/Nearby_Dwarf_Database.html}} (though we take the velocity dispersions for Phoenix and Tucana from \citealt{Kacharov2017} and \citealt{Taibi2020}, respectively.). We also compare to the Milky Way satellites sample assembled from the literature in \citet{Simon2019}. We additionally compare to the homogeneously analyzed outer halo satellites sample of \citet{Munoz2018}. For the latter catalog, we have used their best-fit parameters assuming an exponential density profile in order to better compare to our galaxy morphological fits. In cases where the same galaxy may exist in multiple catalogs, we show values only for the more recent estimate.  We exclude observed galaxies that are more than 1.5~Mpc from the Milky Way, in order to keep their environments comparable to our simulations.

In comparing to our simulations, we assume a virial radius of 300~kpc for both the Milky Way and M31. We exclude all observed dwarf galaxies with half-light radii below 50~pc; this is approximately the radius at which size alone cannot distinguish objects as galaxies versus globular clusters \citep{Simon2019}, and all such galaxies are below the resolution limit of the simulations, so we do not in general expect to be able to reproduce them. For clarity in plot comparisons, we exclude observed properties with large uncertainties (e.g., uncertainty in $M_V$ greater than 5); however, if the uncertainty is missing for $M_V$, we assume it to be 1.

\subsection{Luminosity Functions}

\begin{figure*}[h]
    \epsscale{1.1}
    \plotone{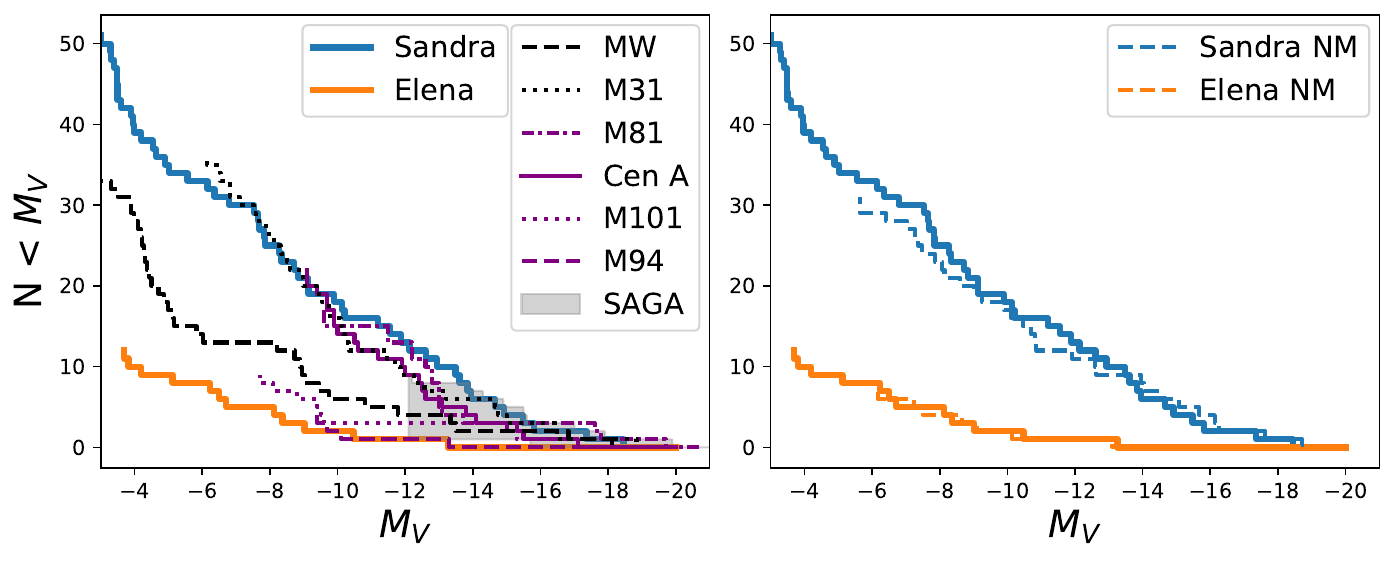}
    \caption{Luminosity functions of the simulated galaxies. In the left panel we compare to the Milky Way and M31, as well as M94 \citep{Smercina2018}, M101 \citep{Bennet2019, Bennet2020}, Centaurus A \citep[Cen A;][]{Crnojevic2019}, and M81 \citep{Chiboucas2013}, down to their completeness limits. We note that the Milky Way luminosity function shown here is almost certainly incomplete below $M_V\sim-8$, with new satellites likely to be found outside the footprints of existing surveys \citep[e.g.][]{Newton2018, Nadler2020}. The SAGA results \citep{Geha2017} are shown as the grey band, representing the full range of luminosity functions from their survey. Sandra and Elena are largely consistent with luminosity functions from the literature, bracketing the range of observed satellite populations. Commensurate with its larger mass, Sandra hosts many more satellites, and is more similar to M31 or the M81 and Cen A groups, while the lower mass Elena is comparable to the sparsely populated M94 system. In the right panel we compare to the Near Mint resolution versions of Sandra and Elena (Sandra NM and Elena NM), which are consistent down to our resolution cutoffs.}
    \label{fig:lf}
\end{figure*}

The left panel of Figure~\ref{fig:lf} shows the satellite luminosity functions of the simulated galaxies. We compare to the sample of \citet{Simon2019} for Milky Way satellites and the updated version of \citet{McConnachie2012} for M31. We also show results from the SAGA survey first release \citep{Geha2017} down to the survey completeness limit, where we represent the full range of luminosity functions as a grey band. Finally, we also include, down to their completeness limits, the luminosity functions of M94 \citep{Smercina2018}, M101 \citep{Bennet2019, Bennet2020}, Centaurus A \citep[Cen A;][]{Crnojevic2019}, and M81\footnote{Both M81 and Cen A are considered small groups, thought to reside in halos more massive than the Milky Way. However, their proximity allows for some of the most complete luminosity functions of host systems close to the Milky Way in mass, so we choose to include them for comparison. The orbital masses of M81 and Cen A have been estimated at $ (4.89\pm1.41)\times10^{12}$~M$_\sun$ and $(6.71\pm2.09)\times10^{12}$~M$_\sun$, respectively \citep{Karachentsev2014}. However, the same authors found that using different methods for determining the mass could lead to mass estimates that are approximately 50\% smaller for these two systems (see their Tables 3 and 5), which would make Sandra's virial mass consistent with M81 and within a factor of 2 of Cen A.} \citep{Chiboucas2013}, where for M81 we have included galaxies within a projected distance of 300~kpc. We find that our simulations fall within the range of observed luminosity functions for this mass range; Elena has fewer satellites than the Milky Way, and is more consistent with M94, while Sandra is more similar to M31 or the more massive M81 and Cen A systems. Given the halo masses of the simulated galaxies, it is unsurprising that Sandra would have significantly more satellites. Elena and Sandra are also in line with results from the SAGA survey, though Elena is among the sparsest systems. Together, Elena and Sandra seem to bracket the observed range of luminosity functions very well.

The right panel of Figure~\ref{fig:lf} shows the satellite luminosity functions of the Near Mint (2x lower force resolution) versions of each simulation, indicated by ``Elena NM'' and ``Sandra NM.'' The luminosity functions are consistent between the Near Mint and Mint resolution simulations. The results are even converged down to the faintest galaxies in the Near Mint sample. For the Near Mint simulations, we applied the same resolution criteria (N$_\mathrm{star}\ge10$ and N$_\mathrm{dark}\ge1000$ at peak halo mass) for inclusion in the sample. It is therefore reassuring that galaxy global properties are converged down to 10 star particles, consistent with \citet{FIRE-2}. Additionally, the Near Mint simulations are in fact able to probe the very brightest UFDs, and while their resolution is lower than the simulations presented in this work, they are still comparable to other high resolution studies of the Milky Way \citep[e.g.,][]{Sawala2016, Grand2017, Garrison-Kimmel2019, Buck2020}.

We expect existing observations of the Milky Way to be largely complete brighter than $M_V\sim-8$, but in the UFD regime a full census is likely to more than double the number of known satellites. By combining the satellite distribution and survey coverage of SDSS and DES along with the radial subhalo distribution from the (dark matter-only) Aquarius simulations \citep{Springel2008}, \citet{Newton2018} predict that there should be ${\sim}40$ galaxies brighter than $M_V=-4$ within 300 kpc of the Milky Way. This indicates that we may be under-producing the faintest galaxies in our simulations, or that adjustments to our halo- and galaxy-finding procedure are necessary. On the other hand, a large fraction of the UFDs discovered in DES are thought to be associated with the Magellanic Clouds \citep[e.g.,][]{Deason2015, Jethwa2016, Sales2017, Kallivayalil2018, Li2018}, the exclusion of which would substantially lower the faint-end luminosity function, alleviating the tension. Since Sandra appears fully consistent with the expectations of \citet{Newton2018}, it is likely that Elena's lower mass and lack of LMC are sufficient to explain any tension.

While we only compare to two simulations, the different luminosity functions strongly suggest a dependence of the total satellite population on the host halo mass \citep[see also][]{Carlsten2020}, in contrast to e.g., \citet{Samuel2020}, who find little correlation. We note, however, that our mass range is a factor of two larger than in their work, which may account for the difference.

\begin{figure*}
    \epsscale{1.1}
    \plotone{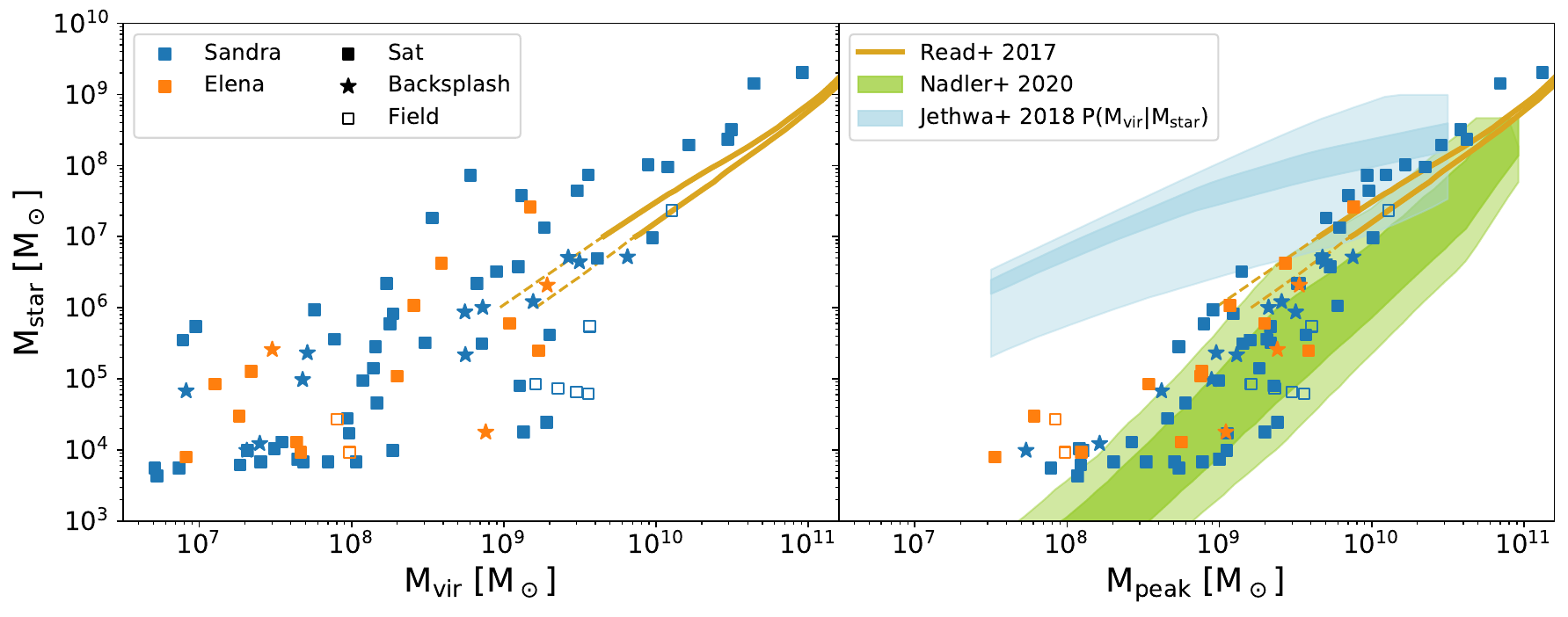}
    \caption{The stellar and halo masses of galaxies in our sample, for both simulations in our suite. Satellite galaxies are shown as filled squares, backsplash galaxies are shown as stars, and field galaxies are shown as empty squares. The left panel shows galaxies' present-day halo masses, while the right panel shows their peak halo masses through time. Both panels show $z=0$ stellar masses. We compare to the stellar mass-halo mass relation inferred in \citet{Nadler2020}, as well as the halo occupation + scatter model of \citet{Jethwa2018}, where the dark (light) bands represent the 68 (95)\% confidence intervals. Both \citet{Nadler2020} and \citet{Jethwa2018} derive their relations via a Bayesian analysis of the Milky Way's observed satellites. We also compare to the relation of \citet{Read2017}, which uses halo masses inferred from HI rotation curves of isolated field dwarf galaxies; the lines enclose the inner 68\% confidence interval, and we use dashed lines to indicate an extrapolation of their relation.}
    \label{fig:smhm}
\end{figure*}

\subsection{Stellar Mass-Halo Mass Relations}

To explore the galaxy-halo connection, in Figure~\ref{fig:smhm} we show the stellar mass-halo mass (SMHM) relation for all galaxies in our sample, along with recent results from the literature for comparison purposes \citep{Read2017, Jethwa2018, Nadler2020}\footnote{We note that these simulations were calibrated to match the relation from \citet{Moster2013}, albeit at higher masses than the dwarfs studied here.}. We note that at low masses, not all halos host galaxy counterparts, and we have chosen to compare to relations that incorporate this fractional occupation. We show both centrals and satellites. The left panel shows stellar mass as a function of present-day virial mass, while the right panel uses M$_{\rm peak}$, the peak halo mass (i.e., before stripping), as done in abundance matching techniques.

As was shown in the satellite luminosity functions of Figure~\ref{fig:lf}, Sandra hosts many more satellites and nearby galaxies than Elena, in approximate proportion to the higher mass of the main halo. However, the SMHM relation appears to be consistent between the two runs. As in previous works \citep[e.g.,][]{Garrison-Kimmel2017, Munshi2017}, we find increasing scatter at lower masses, extending all the way to the UFD regime. Nonetheless, above M$_{\rm peak} \sim 10^{10}$ M$_{\odot}$ our results appear to be consistent with the results of \citet{Read2017}. At lower masses, we largely overlap with the results of \citet{Nadler2020}. Our results appear inconsistent with the relation of \citet{Jethwa2018}, but this is due to our use of their $P(\mathrm{M}_\mathrm{vir}|\mathrm{M}_\mathrm{star})$ relation, whereas their $P(\mathrm{M}_\mathrm{star}|\mathrm{M}_\mathrm{vir})$ is more similar to \citet{Nadler2020}. As discussed by \citet{Jethwa2018}, the former relation is better motivated for inferring halo masses of observed galaxies, in contrast to traditional abundance matching results. Since our SMHM relation is not derived from abundance matching, we choose to compare to $P(\mathrm{M}_\mathrm{vir}|\mathrm{M}_\mathrm{star})$, even though it appears more discrepant to our results.

In the left-hand panel, satellite galaxies exhibit the largest scatter, where tidal interactions with the main halo can lead to preferential mass loss of the dark matter content of halos and drive significant changes in the ratio of stellar mass and halo mass \citep[e.g.][]{Jackson2020}. When using the peak halo mass, the scatter is greatly reduced, though it still increases at low masses. For an in-depth analysis of scatter and more regarding the SMHM relation, we refer the reader to Munshi~et~al.~(in-prep), 
which will update the results of \citet{Munshi2017} using the larger Marvel $+$ DC Justice League sample.

\begin{figure}
    \epsscale{1.1}
    \plotone{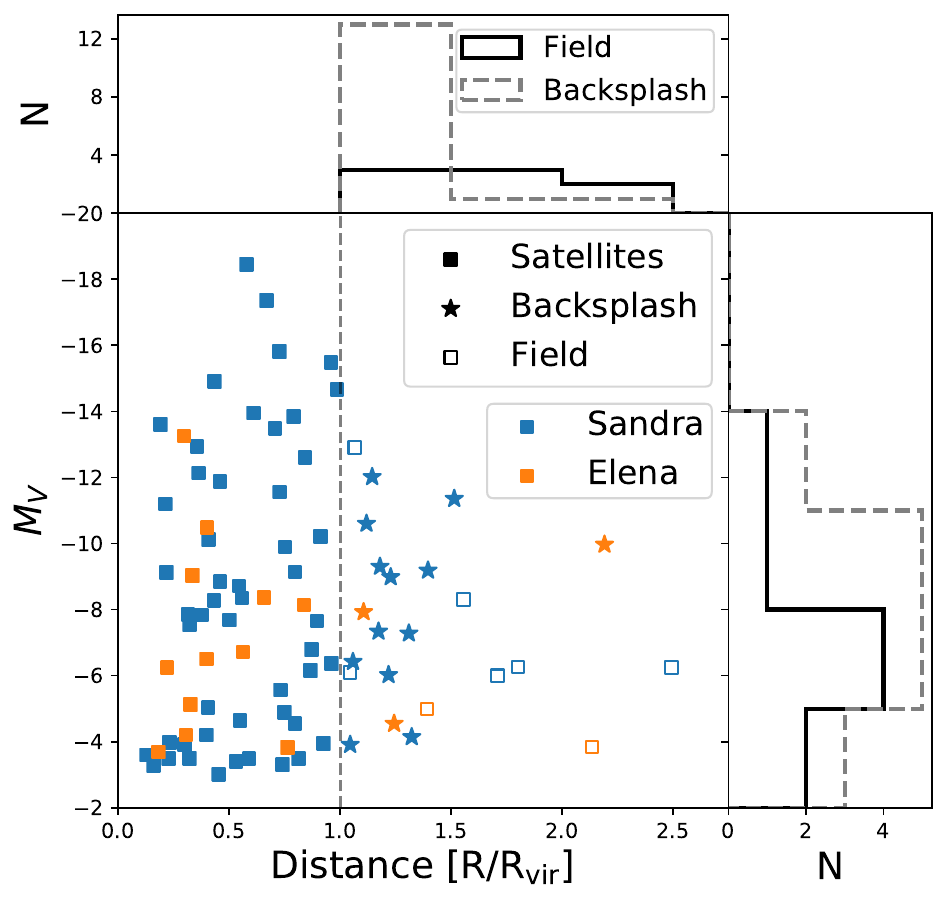}
    \caption{Magnitude vs distance from the central Milky Way for galaxies in both simulations. Satellites are shown as filled squares, field galaxies are empty squares, and backsplash galaxies (galaxies that were previously within the virial radius of the Milky Way) are shown as stars. The vertical dashed line represents the virial radius, for visualization purposes. Marginalized histograms are shown for field (solid black lines) and backsplash (dashed grey lines) galaxies. Just beyond R$_\mathrm{vir}$, most ``field'' galaxies are backsplash galaxies, while beyond ${\sim}1.5$~R$_\mathrm{vir}$ most galaxies have never fallen into the Milky Way. Over half of the galaxies beyond the virial radius are backsplash galaxies, but there is no trend with luminosity.}
    \label{fig:backsplash}
\end{figure}

We note here that a large fraction of galaxies presently in the field have passed within the virial radius of the Milky Way previously. In Figure~\ref{fig:backsplash} we show that in the region just beyond the virial radius, these so-called ``backsplash'' galaxies are the majority of galaxies (as opposed to those that have never had an infall), consistent with previous results \citep[e.g.,][]{Teyssier2012, Buck2019}. Over half of all present-day field galaxies in our sample have had at least one infall within the virial radius of the Milky Way. As we show later in this work, these passages through the Milky Way halo can substantially alter the galaxies' kinematic and structural properties.

\subsection{Galaxy Sizes}\label{sec:structural}

\begin{figure*}
    \epsscale{1.1}
    \plotone{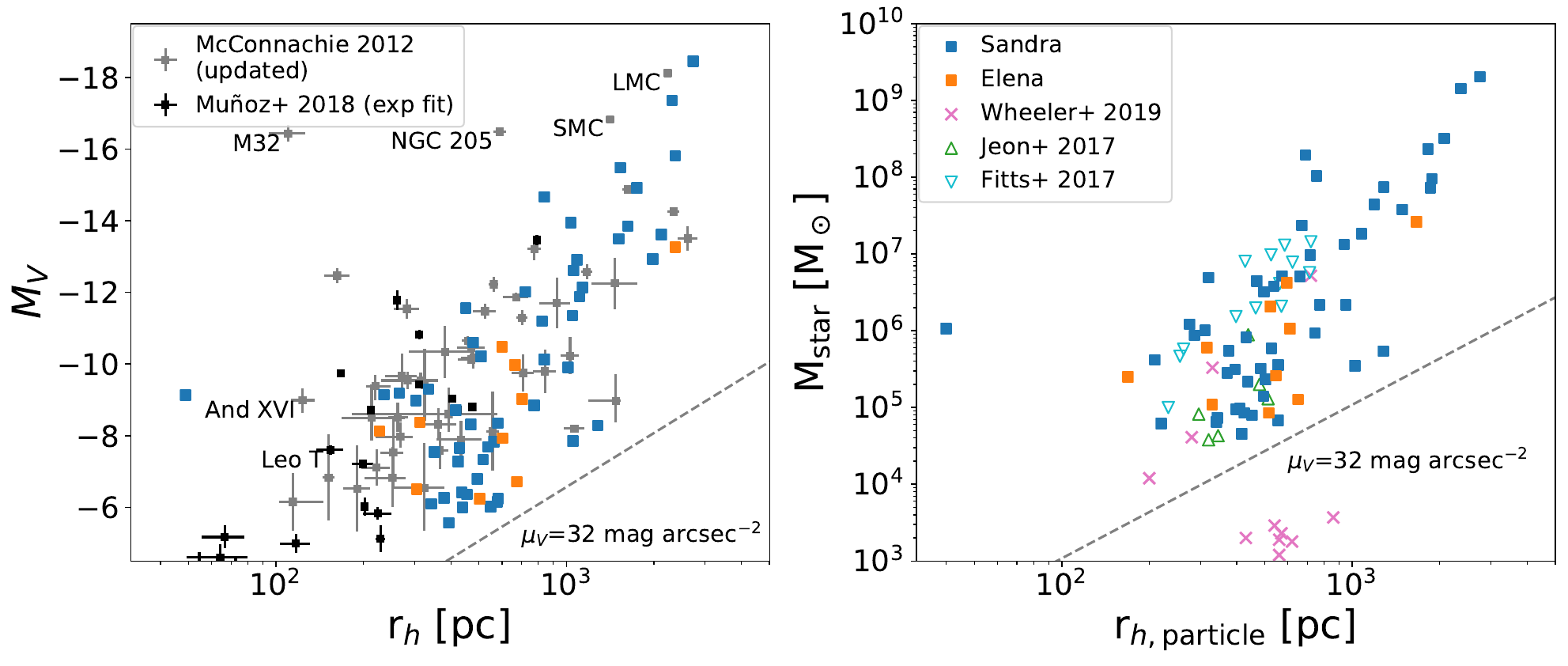}
    \caption{Magnitude versus half-light radius for galaxies with at least 50 star particles. Galaxies are oriented as viewed on the sky from the central simulated Milky Way galaxy. In the left panel, half-light radii are derived via a maximum likelihood estimate of a 2D elliptical exponential profile \citep{Martin2008}. We compare to observed dwarf galaxies in the updated catalog of \citet{McConnachie2012} and the sample from \citet{Munoz2018}; see Section~\ref{sec:obssample} for more detail. We take the values for Crater 2 and the Magellanic Clouds from \citet{Torrealba2016}. In the right panel, we instead show stellar mass, and (circular) half-light radii as derived via direct summation of the particle luminosities. We compare to the simulated samples of \citet{Jeon2017}, \citet{Fitts2017}, and \citet{Wheeler2019}. The dashed lines represent a constant surface brightness of $\mu_V=32$~mag~arcsec$^{-2}$, roughly the limit of co-added Vera Rubin Observatory's LSST. We expect essentially all galaxies near the Milky Way to be observable by LSST. Above $r_h\approx200$~pc, the simulated galaxies reproduce the full range of observations and scatter in the size-luminosity plane.}
    \label{fig:sizes}
\end{figure*}

To better compare the structural parameters of the galaxies to observations, we use maximum likelihood estimation to find the best fitting parameters for a 2D elliptical exponential density profile\footnote{These fits actually find half-density radii rather than half-light radii. As a check, we have fit to images of the V-band luminosity for these systems and found no systematic differences in the parameter estimates.}; for more detail, see \citet{Martin2008}. The density profile has the following functional form:
\begin{equation}
    \Sigma(r) = \Sigma_0 e^{r/r_e},
\end{equation}
where $r_e$ is the scale radius (with a half-light radius given by \mbox{$r_h=1.68r_e$)}, $\Sigma_0$ is the central density, and $r$ is the elliptical radius given by
\begin{equation}
    r = \left\{\left[\frac{1}{1-\epsilon}\left(X\cos\theta - Y\sin\theta \right)\right]^2 + \left(X\sin\theta+Y\cos\theta\right)^2\right\}^{1/2},\label{eq:elliptical_radius}
\end{equation}
where $\epsilon$ is the ellipticity\footnote{The ellipticity is defined as $\epsilon=1-b/a$, where $a$ and $b$ are scale lengths along the major and minor axes, respectively.} and $\theta$ is the angular offset of the ellipse from the vertical. We additionally simultaneously fit for the centroid $(x_0, y_0)$ of the ellipse, such that $X_i=x_i-x_0$ and $Y_i=y_i-y_0$.

Figure~\ref{fig:sizes} shows the size-luminosity relationship as viewed on the sky from the central simulated Milky Way galaxy. In the left panel we compare to observational data from \citet{McConnachie2012} and \citet{Munoz2018}, and show half-light radii calculated for the simulated galaxies as described above. We also show the line of constant mean surface brightness, $\mu_V=32$~mag~arcsec$^{-2}$, which is approximately the limiting surface brightness of the Vera Rubin Observatory's co-added Legacy Survey of Space and Time \citep{LSST2009}. In the right panel we compare to the results from several cosmological simulations of field UFDs \citep{Jeon2017, Fitts2017, Wheeler2019}. For this comparison, we use the projected, circular half-light radius as calculated via direct summation of the particle luminosities, and show stellar mass in lieu of magnitude. The half-light radii of \citet{Fitts2017} are 3D, so we multiply the listed values by $3/4$ to approximate the 2D projected half-light radii \citep{Wolf2010}. For the line of constant surface brightness, we have assumed a stellar mass-to-light ratio of 3, consistent with the faintest and oldest galaxies.

Across most of the sample, including in the ultra-faint range, we reproduce observed galaxy sizes\footnote{We have ensured these results are robust by fitting along many random lines of sight. As in \citet{El-Badry2016}, we find that half-light radii vary typically by no more than ${\sim}10$\%}. At a given magnitude, the galaxies span a range of sizes, including some diffuse and some relatively compact galaxies. Our most compact galaxies are generally about as bright and slightly larger than Andromeda XVI or Leo T; reproducing these galaxies has been a challenge in some previous works \citep[e.g.,][]{Revaz2018, Garrison-Kimmel2019}. There is one additional galaxy with $M_V\sim-9$ and $r_h=40$~pc that is by a wide margin the most compact simulated dwarf. This galaxy seems to be a faint analog of ultra-compact dwarf (UCD) galaxies, and understanding its evolution may shed light on the origin of UCD galaxies. Our sample includes just one of these galaxies; it is possible that a larger simulation suite---or a more massive central galaxy---would include more of them, or even brighter ones. In Section~\ref{sec:ucd} we discuss this compact galaxy in more detail, including its evolutionary history.

The UCD analog represents a step forward in modeling compact galaxies. However, we produce no other galaxies with half-light radii below ${\approx}200$~pc\footnote{We note that the smallest non-UCD galaxy has a half-light radii just above 200~pc when calculated using the elliptical fit, and just under 200~pc when circular symmetry is imposed.}, as might be expected given that this is just a few times our force softening. Indeed, at $M_V\sim-6$, most of our galaxies are larger than those observed to date (though still above the 32~mag~arcsec${}^{-2}$ line); given that they are just outside the observed range, they may indicate that there are dimmer UFD galaxies yet to be discovered. The absence of additional compact galaxies, as well as the lack of compact bright galaxies like M32, is a manifestation of the ``diversity'' problem \citep{Oman2015}, for which there is as yet no accepted solution\footnote{In non-cosmological contexts, however, simulations have successfully reproduced M32-like galaxies \citep{Du2019}}. It is possible that we lack the ability to resolve such dense galaxies, or that spurious dynamical heating from 2-body interactions systematically increases half-light radii by $z=0$ \citep{Revaz2018, Ludlow2019a}.

Nonetheless, above $r_h\approx200$~pc, we do produce the \textit{entire} observed range of sizes and luminosities. Prior works, while overlapping with these simulations in the size-luminosity plane, show less scatter, independent of sample size. To quantify the scatter, we performed the following procedure: we derived best-fit lines in the $M_V-\log\,r_h$ and $\log\,\mathrm{M}_\mathrm{star}-\log\,r_h$ planes for our simulations, as well as for each prior observational and simulation work shown in Figure~\ref{fig:sizes}, for all galaxies with $r_h>100$~pc. We then calculated the sample standard deviations of the residuals to the fits. We found that the scatter in our simulations was about twice that of \citet{Fitts2017} and \citet{Jeon2017}, and consistent with that of the observations ($\sigma_{M_V}\approx2$, $\sigma_{\log\,r_h}\approx0.2$), regardless of whether we calculated the fit and residuals along the size axis or mass/luminosity axis. To emulate the smaller sample sizes of the prior simulation works, we also repeated the procedure for $10^4$ random samples of 6 galaxies each from our simulations, and found our scatter to be higher than theirs in 90\% of cases. Of the field dwarf simulations we compare to, only \citet{Wheeler2019} had comparable scatter to that of our simulations and the observations.

Importantly, none of our galaxies have mean central surface brightnesses dimmer than 32~mag~arcsec$^{-2}$. We therefore expect essentially all nearby galaxies to be observable by the Vera Rubin Observatory's co-added LSST, with no galaxies too diffuse to detect in the UFD galaxy range, at least down to the luminosity limit probed here. In the mass range that we are able to resolve, \citet{Wheeler2019} find similar results. However, below $10^4$~M$_\sun$ in stellar mass, all of the UFDs in \citet{Wheeler2019} are more diffuse than those observed. We discuss numerical differences, including feedback implementations and resolution, in Section~\ref{sec:previouswork}.

\subsection{Kinematics}\label{sec:kin}

\begin{figure}
    \epsscale{1.15}
    \plotone{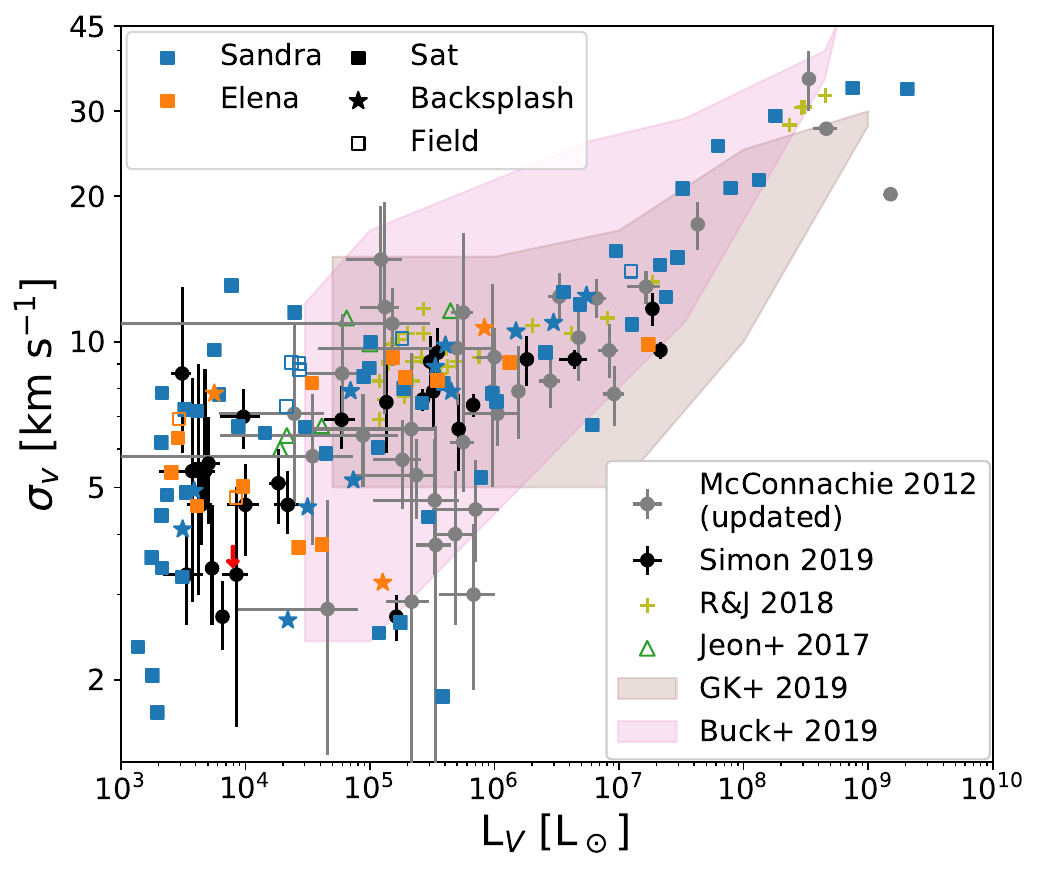}
    \caption{Line-of-sight velocity dispersion of the simulated dwarf galaxies as a function of luminosity for both simulations (Sandra and Elena). We show satellites as filled squares, backsplash galaxies as stars, and field galaxies as empty squares. We compare to observed dwarfs using the compilation of \citet{Simon2019} and the updated version of \citet{McConnachie2012}. We also compare to previous simulations, including the field dwarf galaxies of \citet{Jeon2017} and \citet{Revaz2018}, the Milky Way simulations of \citet{Buck2019}, and the Milky Way and Local Group simulations of \citet{Garrison-Kimmel2019}. For clarity, for the latter two simulations we show bands that approximate the full range of their results. For \citet{Jeon2017}, \citet{Buck2019}, and \citet{Garrison-Kimmel2019}, we assume a stellar mass-to-light ratio of 2 for galaxies fainter than $10^6$~$L_\sun$, and a ratio of 1 otherwise. While the prior field simulations tend to have higher velocity dispersion, our simulations reproduce the full range. The dynamically coldest systems ($\sigma_v\lesssim5$~km/s) have all been severely tidally stripped; see Figure~\ref{fig:sigma_massloss}.}
    \label{fig:kin}
\end{figure}

Figure~\ref{fig:kin} shows the line-of-sight velocity dispersions for galaxies in the simulations, separated by environment. We compare to observations from \citet{McConnachie2012} and \citet{Simon2019}, as well as previous simulations of field dwarfs \citep{Jeon2017, Revaz2018}. We also compare to the FIRE-2 simulations of \citet{Garrison-Kimmel2019}, who investigated dwarf galaxies with M$_\mathrm{star}>10^5$~M$_\sun$ in a suite of Milky Way-like and Local Group-like simulations, and the NIHAO simulations of \citet{Buck2019}. Both the latter simulations have large samples, so we show their results as shaded bands that approximate their full range of values.

The kinematics of the simulated DC Justice League galaxies reproduce those of observed galaxies. For luminosities $L_V < 10^6$ L$_{\odot}$, galaxies show a large scatter, but that scatter is relatively constant down to the faintest galaxies, such that galaxies with $L_V \sim 10^4$ L$_{\odot}$ and galaxies with $L_V \sim 10^6$ L$_{\odot}$ appear to span a similar range of masses/kinematics, just as seen in the observations.  However, for $L_V > 10^6$ L$_{\odot}$ galaxies have higher velocity dispersions. In our simulations, galaxies with M$_\mathrm{star}\gtrsim10^{7}$~M$_\sun$ are able to form cored dark matter density profiles.  Dark matter cores allow for more substantial tidal stripping that can also lead to smaller velocity dispersion, particularly for those satellites with small orbital pericenters \citep[e.g.,][]{Brooks2014}.

Unlike most of the prior simulations we compare to in Figure~\ref{fig:kin}, we produce galaxies with $\sigma_v<5$~km/s, consistent with many observed galaxies. This can largely be explained by the effects of environment: all of the low-$\sigma_v$ galaxies in our simulations are either current satellites or backsplash galaxies. \citet{Jeon2017} and \citet{Revaz2018} simulated only isolated field environments. Correspondingly, their galaxies are generally consistent with the galaxies from our simulations that have the highest dispersions at a given luminosity.

\begin{figure}
    \epsscale{1.15}
    \plotone{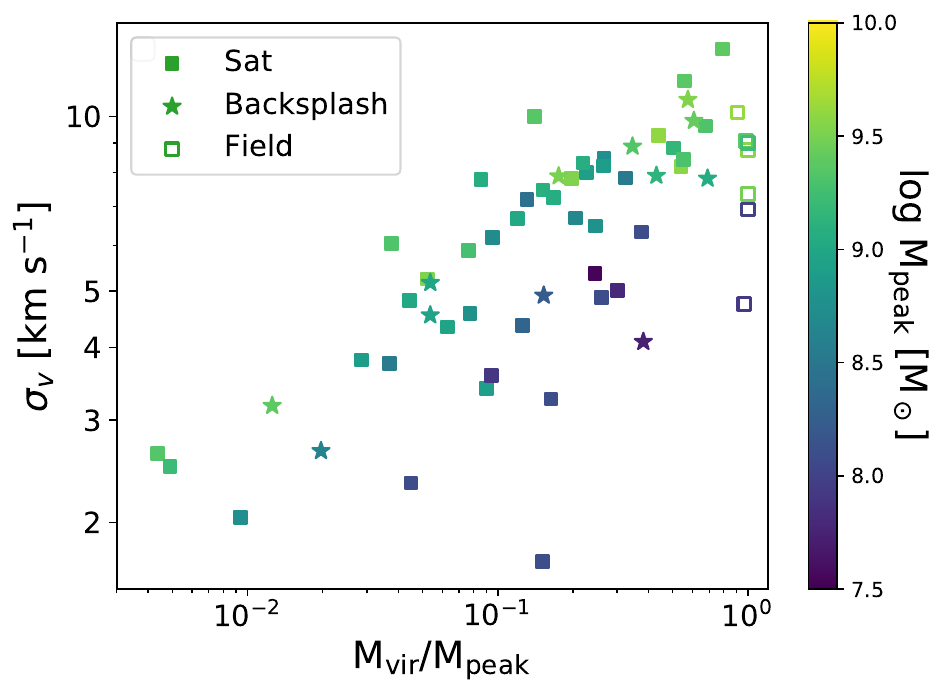}
    \caption{Velocity dispersion as a function of the fraction of mass remaining from peak halo mass for all galaxies in the simulations with $L_V<10^6\,L_\sun$. Present-day satellites are marked with filled squares, backsplash galaxies are marked with stars, and field galaxies are marked with empty squares. There is a tight correlation between velocity dispersion and tidal stripping: the dynamically coldest halos at a given M$_\mathrm{peak}$ have experienced the most mass loss, regardless of present-day location, while less massive halos tend to be intrinsically dynamically colder. All galaxies with $\sigma_v<5$~km/s and $L_V>10^4$~$L_\sun$ (all of which have M$_\mathrm{peak}>10^{8.5}$~M$_\sun$) have lost at least 90\% of their mass.}
    \label{fig:sigma_massloss}
\end{figure}

Previous work has consistently shown that severe tidal stripping in the presence of a massive host can lead to lower velocity dispersions \citep[e.g.,][]{Penarrubia2008, Brooks2014, Errani2015, Frings2017, Fattahi2018, Buck2019, Maccio2020}, explaining the difference between environments. \citet{Penarrubia2008} found that galaxy structure and kinematics tend to follow evolutionary tracks that depend mainly on how much total mass has been lost, with $\sigma_v$ decreasing monotonically as mass loss increases. In Figure~\ref{fig:sigma_massloss} we show the same is true in our simulations. The figure shows the present-day velocity dispersion for all galaxies with $L_V<10^6\,L_\sun$ as a function of their (total) mass loss from peak. We separate by environment, though $\sigma_v$ appears to only depend on mass loss, not present-day location.

Figure~\ref{fig:sigma_massloss} directly shows the importance of simulating faint galaxies in the context of the Milky Way environment. All but one of the galaxies with $\sigma_v<5$~km/s have lost most of their mass due to tidal stripping, and all galaxies with both $L_V>10^4$~$L_\sun$ and $\sigma_v<5$~km/s (all of which have M$_\mathrm{peak}>10^{8.5}$~M$_\sun$) have lost at least 90\% of their mass. Importantly, even some backsplash galaxies that today are in the field have experienced severe tidal stripping. The figure also shows a mass-dependent trend in velocity dispersion. While tidal stripping leads to lower dispersions in all halos, galaxies in smaller halos are systematically dynamically colder; the coldest field galaxy has $\sigma_v=4.8$~km/s, despite having lost less than 4\% of its mass.  

Tidal effects explain why the field simulations in Figure~\ref{fig:kin} do not contain low-$\sigma_v$ galaxies. \citet{Buck2019} are the only other set of simulations that produce galaxies with velocity dispersions below 5~km/s. We note that their simulations were run with a modified version of \textsc{Gasoline}, which uses the same hydrodynamics solver upon which \textsc{ChaNGa} is based, as well as similar feedback recipes. However, their simulations were run at lower resolution and with a density-based star formation scheme. \textsc{ChaNGa}'s H$_2$-based star formation scheme leads to low velocity dispersions at birth \citep{Bird2020}, and may be necessary in order to reproduce the galaxies with $L_V<10^4$~$L_\sun$ and $\sigma_v<5$~km/s, which have generally been less severely tidally stripped than their more luminous counterparts. It is interesting that the Milky Way and Local Group simulations of \citet{Garrison-Kimmel2019} also do not reproduce the lowest $\sigma_v$ galaxies, despite capturing the same environmental effects as our simulations, and despite adopting a star formation prescription that should also capture the high densities and low temperatures of gas forming in H$_2$.  They discuss several possible explanations, including insufficient resolution, N-body dynamical heating, or spurious (numerical) subhalo disruption \citep{vandenBosch2018}. 

\subsection{Mass-to-Light Ratios}

\begin{figure}
    \epsscale{1.1}
    \plotone{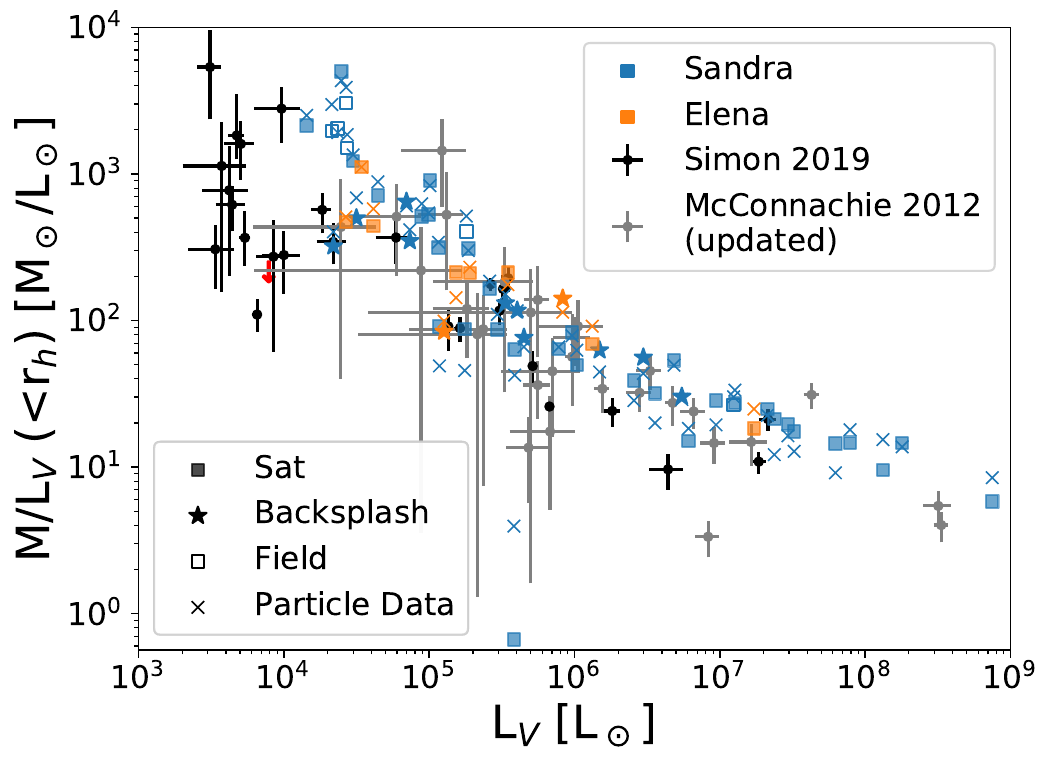}
    \caption{Mass-to-light ratios within the half-light radius for galaxies with at least 50 star particles. We show the values as derived using the \citet{Wolf2010} mass estimator (equation~\ref{eq:wolf2010}), separated into present-day field, satellite, and backsplash populations. For all galaxies, we also show the mass-to-light ratios as derived from the simulation particle data; the two methods are generally consistent with each other, even at higher luminosities where the systems are no longer dispersion-supported. We compare to observed dwarf galaxies, whose values have all been derived using the \citet{Wolf2010} relation. The case where only an upper limit exists on the observations is shown as a downward red arrow. Across the whole range of luminosities, the simulated galaxies match the observed relation. In the faint end, simulated galaxies with higher mass-to-light ratios tend to be field galaxies, unlike the observational data. We note that above ${\sim}10^7 L_\sun$, galaxies transition to rotation support, so the inferred mass-to-light ratios should be treated with caution.}
    \label{fig:mlratio}
\end{figure}

We show in Figure~\ref{fig:mlratio} the mass-to-light ratio within the half-light radius as a function of $V$-band luminosity. Previous studies have shown that enclosed mass is robustly estimated within the observed half-light radius \citep[e.g.,][]{Walker2009, Wolf2010}. For both the observed and simulated galaxies, we therefore use the methodology of \citet{Wolf2010} to calculate the dynamical mass within the half-light radius, or
\begin{equation}
    M_{1/2}\approx930\,\left(\frac{\sigma_v^2}{\mathrm{km}^2\,\mathrm{s}^{-2}}\right)\,\left(\frac{r_h}{\mathrm{pc}}\right)\,\mathrm{M}_\sun,\label{eq:wolf2010}
\end{equation}
where $\sigma_v$ is the line-of-sight velocity dispersion (as seen from the Milky Way) and $r_h$ is the half-light radius. When calculating the mass-to-light ratios, we find the mass within the ``circularized'' half-light radius $r_h\sqrt{1-\epsilon}$ \citep{Sanders2016}, where $\epsilon$ is the ellipticity of the system as seen from the Milky Way. For comparison, we also plot the mass-to-light ratios calculated by directly summing the particle data enclosed within the half-light radius.

The results in Figure~\ref{fig:mlratio} match the observational data, whose masses are also derived using equation~\ref{eq:wolf2010}. The simulations reproduce the general trend, as well as the scatter, with the faintest systems being dominated by dark matter. For the most part, the mass-to-light ratios derived from equation~\ref{eq:wolf2010} are close to the true values derived from particle data, indicating the general robustness of the \citet{Wolf2010} estimator (see also \citealt{Campbell2017, Gonzalez-Samaniego2017}). The outlier to this trend, with $L_V\sim10^{5.5}$~$L_\sun$ and a particle-derived mass-to-light ratio $<10$, has a ${\sim}0.5$~dex lower ratio when inferred from the velocity dispersion. This galaxy corresponds to the compact system (see Figure~\ref{fig:sizes}) with limited dark matter content; it is seemingly an ultra-compact dwarf analog, which we discuss in more detail in Section~\ref{sec:ucd}.

A few of the simulated galaxies with $L_V<10^{4.5}\,L_\sun$ are more dark matter-dominated than observed systems in the same luminosity range. These galaxies are mostly field galaxies that have never had an infall---in other words, they have never been substantially tidally stripped. As shown in the previous section, these are the galaxies that are dynamically hottest in this luminosity range. On the other hand, all of the \textit{observed} galaxies in this range are either satellites of the Milky Way or Andromeda; the observations may therefore be biased toward more heavily stripped halos, and subsequently lower mass-to-light ratios.

We additionally note that in Figure~\ref{fig:mlratio}, galaxies with $L_V\gtrsim10^7$~$L_\sun$ transition from primarily dispersion-supported to rotation-supported galaxies, and so equation~\ref{eq:wolf2010} is no longer valid. Despite this, for the simulated galaxies the estimator remains consistent with the particle data across the entire range of luminosities, demonstrating its robustness. Nonetheless the mass-to-light ratios of the brightest observed galaxies in the figure should be treated with caution.

\subsection{Metallicities}

\begin{figure*}
    \epsscale{1.1}
    \plotone{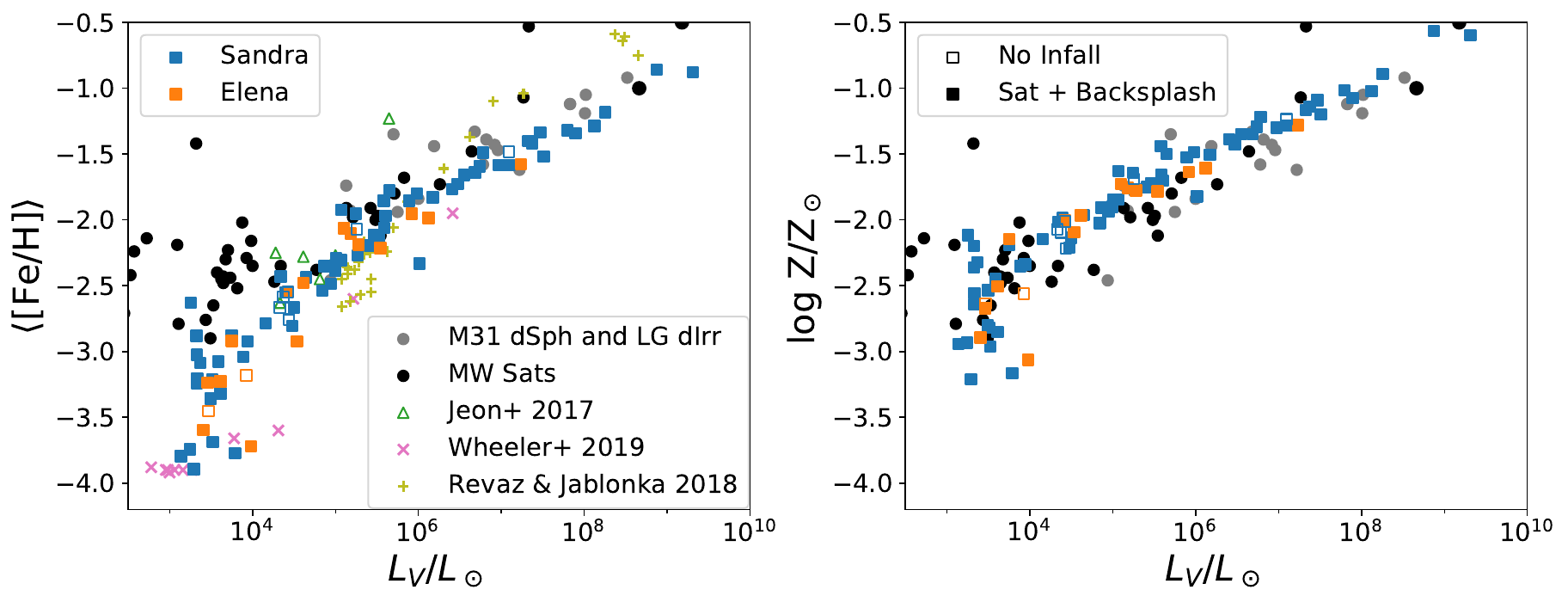}
    \caption{Stellar metallicities of the simulated galaxies as a function of luminosity. \textit{Left}: metallicities are calculated as the mean [Fe/H]. Satellite and backsplash galaxies are shown as filled squares, while galaxies that have never had an infall are shown as empty squares. Data of observed satellite galaxies are taken from the compilation by \citet{Simon2019}, except for the Magellanic Clouds, which are taken from \citealt{McConnachie2012}, while non-satellite dwarf irregular galaxies and Andromeda satellite dwarf Spheroidals are from \citet{Kirby2013b}. We also show simulated field dwarf galaxies from \citet{Jeon2017}, \citet{Revaz2018}, and \citet{Wheeler2019}. Since the simulations of \citet{Wheeler2019} apply a metallicity floor of \mbox{[Fe/H] $=-4$}, we have applied the same floor to any star particles with a lower metallicity in our simulations. There is broad agreement between the observations and the simulations presented in this work for galaxies with $L_V\gtrsim10^4$~$L_\sun$. For galaxies below 10$^4$~$L_\sun$, simulated metallicities appear to be lower than observations, but substantially more metal rich than in \citet{Wheeler2019}. \textit{Right}: metallicities are calculated from total metals, also applying a floor of $\log Z/Z_\sun=-4$. Symbols are the same as the left panel, except we no longer compare to prior simulations. Compared to the left panel, agreement is improved even further, especially for the faintest galaxies. This agreement indicates that the simulated UFDs are able to retain metals in the ISM, but may be under-producing iron.}
    \label{fig:mzr}
\end{figure*}

Observations of Local Group galaxies follow a universal relationship between stellar mass (or luminosity) and stellar metallicity, across orders of magnitude in mass and across various morphologies \citep[e.g.,][]{Kirby2013a, Kirby2019}. At higher masses, numerous groups are now able to reproduce these trends \citep[e.g.,][]{Brooks2007, Ma2016, DeRossi2017, Christensen2018, Torrey2019}. At lower masses, however, most simulations produce galaxies with stellar metallicities below those observed \citep[e.g.,][]{Maccio2017, Revaz2018, Wheeler2019}. Part of the challenge stems from the need to both ensure inefficient star formation via feedback, while also retaining metals in the interstellar medium to be incorporated in subsequent generations of stars. Multiple explanations have been offered to explain the too-low stellar metallicities, including pre-enrichment from Population III (Pop III) stars or varying IMF yields \citep[e.g.,][]{Revaz2018, Wheeler2019}, insufficient time resolution \citep{Maccio2017}, pre-enrichment from the more massive host galaxy \citep{Wheeler2019}, or too-efficient feedback \citep{Agertz2020}.

The left panel of Figure~\ref{fig:mzr} shows the luminosity-metallicity relationship for all galaxies in the sample. To maximally separate out the effects of environment, we plot satellite \textit{and} backsplash galaxies with filled squares while showing galaxies that have never had an infall as empty squares. We compare to observed Milky Way satellite galaxies \citep{Simon2019, McConnachie2012}, as well as Local Group dIrrs and M31 satellites \citep{Kirby2013a}. We also compare to simulations of field dwarf galaxies from the literature \citep{Jeon2017, Revaz2018, Wheeler2019}. When calculating the mean, we assign a metallicity of \mbox{[Fe/H] $=-4$} to any star particles with \mbox{[Fe/H] $<-4$}. This allows our results to be more directly comparable to those of \citet{Wheeler2019}, who impose this metallicity floor in the simulations themselves.

Galaxies with $L_V\gtrsim10^4$~$L_\sun$ are consistent with observations across the full luminosity range, and there are no systematic differences across environment in Figure~\ref{fig:mzr}, in agreement with the observed data. While the metallicities might be slightly low (but see below), their slope is consistent with observations. In fainter galaxies, with $L_V<10^4$~$L_\sun$, the simulations are less successful at reproducing the observations, with a few of the simulated UFDs having stellar metallicities that are largely unenriched. However, most of the UFDs---even those with as few as 10 star particles---have experienced some level of cumulative chemical enrichment that brings their metallicity above the floor. The more metal-rich UFDs are fully consistent with observed galaxies. This is in contrast to the results of \citet{Wheeler2019}, who found no metal enrichment at all in galaxies with M$_\mathrm{star}<10^4$~M$_\sun$, despite resolving these galaxies with $>100$ star particles. Only at M$_\mathrm{star}>10^5$~M$_\sun$ do their galaxies approach the observed relation.

The discrepant results between this work and \citet{Wheeler2019} may reflect the different feedback implementations between our simulations and the FIRE-2 simulations. Work by \citet{Agertz2020} showed that metallicity is highly sensitive to feedback implementation; we discuss this further in Section~\ref{sec:previouswork}. \citet{Wheeler2019} suggest that a lack of Pop III or environmental pre-enrichment may account for their low simulated metallicities, but contributions from pre-enrichment may be insufficient: while highly uncertain, Pop III yields were likely iron-deficient \citep[e.g.,][]{Iwamoto2005, Ishigaki2014}, and even assuming solar abundance in the yields is unlikely to raise the simulated metallicities to observed values \citep{Agertz2020}. Pre-enrichment from a more massive host is also unlikely, since as we show in the next section, these galaxies tend to quench long before they approach the Milky Way. This is reflected in Figure~\ref{fig:mzr}, which shows that galaxies that have never had an infall in our simulations have comparable metallicities to satellite and backsplash galaxies.

Similarly, Pop III or environmental pre-enrichment likely does not account for the low [Fe/H] of galaxies with $L_V\lesssim10^4$~$L_\sun$ in our simulations. Instead, the timing of star formation may be more important. The right panel of Figure~\ref{fig:mzr} shows the luminosity-metallicity relationship again, but instead uses total stellar metallicity rather than [Fe/H]. The galaxies across the entire luminosity range---including the faintest galaxies---are consistent with the observed data. None of the UFDs are at or near the metallicity floor. This suggests the galaxies are successfully retaining metals in the ISM. 

If the galaxies are both producing and retaining enough metals to enrich to the observed luminosity-metallicity relationship, then the low [Fe/H] in the faintest galaxies implies that they are simply underproducing iron relative to oxygen. Iron is produced predominantly in Type Ia explosions, which occur in our simulations on ${\sim}$Gyr timescales \citep{Raiteri1996}. It is likely, therefore, that star formation is stopping too soon relative to Type Ia delay times. One possibility is that the duration of star formation is too short in the UFDs. Another is that the timescale for Type Ia supernovae is too long, and that we need to include models for ``prompt'' Type Ia supernovae, occurring on ${\sim}100$~Myr timescales \citep{Mannucci2006, Maoz2012}. Finally, Pop III stars may pre-enrich galaxies to a higher [Fe/H] floor, though this seems unlikely as Pop III yields were likely iron-poor \citep[e.g.,][]{Iwamoto2005}.

\section{The Quenching of the Ultra-faints}\label{sec:quenching}

\begin{figure*}
    \epsscale{1.1}
    \plotone{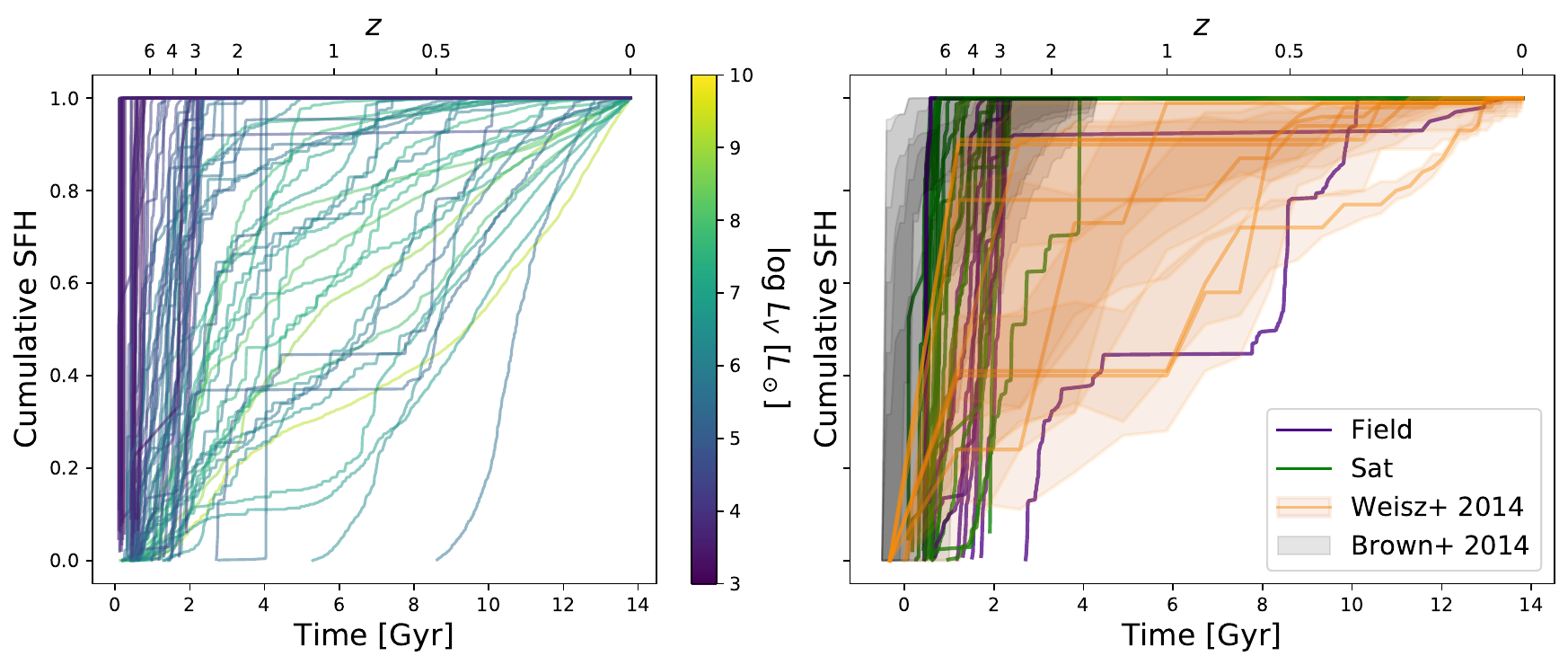}
    \caption{Cumulative star formation histories of galaxies in our sample. The left panel shows all galaxies, colored by their $V$-band luminosity. The dwarf galaxies display a wide array of SFHs, with a general trend that more massive galaxies form their mass later. The right panel shows only UFD galaxies, colored by their present-day environment (satellite or field). For clarity, in the right panel we do not plot simulated UFDs with star formation lasting less than 100~Myr. We also show observed UFD star formation histories derived from color-magnitude diagrams. For \citet{Weisz2014a} we show their best-fit SFH as orange lines and their uncertainty as orange bands, while for \citet{Brown2014} we show as grey bands the full statistical uncertainty range of their cumulative SFHs as derived from their 2-burst model. Our UFDs have generally quick star formation, with most of them quenching by $z\sim3$. The star formation histories are consistent with those from \citet{Brown2014}, but somewhat inconsistent with those of \citet{Weisz2014a}, who find later star formation in some UFDs. However, our results are largely consistent within their uncertainties. We note that both \citet{Brown2014} and \citet{Weisz2014a} use isochrones older than the age of the Universe, and the latter sets the cumulative SFH to 0 at $\log(t) = 10.15$~Gyr; we have made no correction for this, which is why their SFHs appear to start in many cases at $t<0$.}
    \label{fig:csfh}
\end{figure*}

Of the UFDs with resolved star formation histories and ages, most appear to have formed the bulk of their stars early on \citep{Okamoto2012, Brown2014, Weisz2014a, Skillman2017}. Such early quenching is consistent with UFDs being fossils of reionization \citep{Bovill2009}. However, all observed UFDs with constrained SFHs are satellites of the Milky Way or M31 (with the exception of Leo T, which is in the field but may be a backsplash galaxy; see, e.g., \citealt{Blana2020}), so it is difficult to rule out quenching due to interactions with the Milky Way. We compare both satellites of the Milky Way and near-field UFD galaxies in the same simulation, and use their orbital histories to show that feedback from reionization and/or supernovae is the dominant quenching mechanism.

The left panel of Figure~\ref{fig:csfh} shows the cumulative (fractional) star formation histories of all galaxies in the sample, color-coded by $V$-band luminosity. SFHs are calculated from the star particles remaining in the galaxy at $z=0$; stars that may have formed in a galaxy but been tidally stripped are not included. The galaxies exhibit a range of SFHs across all luminosities. On average, however, more massive galaxies form the bulk of their mass later than smaller galaxies. Several galaxies display ``gaps'' in their SFHs where previously quenched galaxies restart their star formation, similar to the phenomenon described in \citet{Wright2019}. There also exist several galaxies that have delayed-onset star formation, with the first star formation starting well after the end of reionization. It is unknown whether any observed galaxies have such late star formation; these will be the subject of future work.  Generally, however, most galaxies begin their star formation before $z\sim6$, and form the majority of their mass by $z\approx2$.

The right panel of Figure~\ref{fig:csfh} includes only galaxies that are in the UFD range, along with the star formation histories of observed UFDs derived from color-magnitude diagarams \citep{Weisz2014a, Brown2014}. For clarity, we plot only galaxies whose star formation lasts at least 100~Myr\footnote{36\% of the UFDs have star formation lasting less than 100~Myr. It is unclear whether any real galaxies have star formation lasting less than 100~Myr; though many UFDs are consistent with exactly single-age populations \citep{Brown2014}, the uncertainty in stellar ages is well above this timescale at ${\sim}1$~Gyr.}; the galaxies with $<100$~Myr SFHs predominantly form as single-age populations within the first 500~Myr after the Big Bang ($z\gtrsim10$). The simulated UFDs are color-coded by their environment (either satellite of the Milky Way or  near-field galaxy). All of the observed UFDs with star formation histories, on the other hand, are satellites of either the Milky Way or Andromeda, with the exception of Leo T. In this luminosity range, most of the simulated galaxies quench by $t\sim3$~Gyr, regardless of whether or not the UFDs are satellites (one of the quenched UFDs restarts its star formation again at later times, which we discuss further in Section~\ref{sec:gasrichufds}). The lack of environmental dependence suggests that quenching is caused by reionization and/or supernova feedback, rather than environmental effects.

Compared to the observations in the right panel of Figure~\ref{fig:csfh}, the simulated UFDs appear to quench faster, with less extended star formation. However, within the total uncertainties of \citet{Weisz2014a}, several of the observed UFDs are consistent with forming all of their stars before $z=3$, as in the simulations. Additionally, the inferred star formation later than $z\sim3$ may be so slight that it would be difficult for the simulations to capture it given the resolution of the star particles. Finally, while there is a slight tension between the simulated SFHs and those of \citet{Weisz2014a}, our results are consistent with the SFHs of the 6 UFDs studied in \citet{Brown2014}\footnote{Three of the UFDs studied in \cite{Brown2014} also have star formation histories from \citet{Weisz2014a}. Two of the three galaxy star formation histories are consistent between the two studies, but the SFH for Canes Venatici II (CVn II) is discrepant for as-yet unknown reasons.}. They found that all the UFDs in their sample formed 80\% of their stars by $z\sim6$ and 100\% by $z\sim3$, as in these simulations.

There is also one late-forming UFD in the right panel of Figure~\ref{fig:csfh}; this is a near-field galaxy just beyond the virial radius of Elena with a $V$-band magnitude of $-7.9$. Unlike the other UFDs in the sample, it began star formation well after reionization and undergoes a different evolution, which we discuss as a case study in Section~\ref{sec:lateformer}.

\begin{figure}
    \epsscale{1.08}
    \plotone{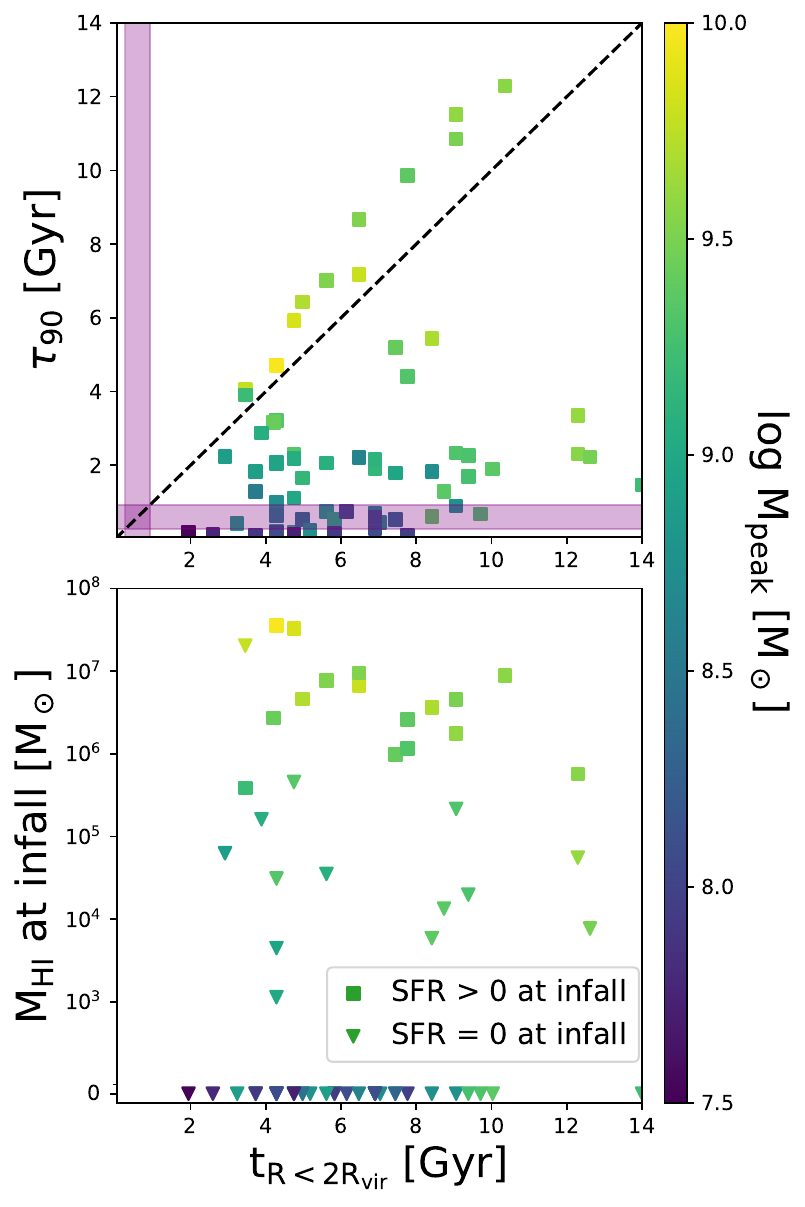}
    \caption{\textit{Top}: Quenching time $\tau_{90}$ vs infall time (to 2~R$_\mathrm{vir}$); galaxies that have not had an infall to this radius are placed on the right-most edge of the figure at 14 Gyr. Galaxies are colored by their peak halo mass. The purple bands show $z=15-6$, which is when reionization occurs in our simulations. The dashed line indicates the one-to-one line. There are two different populations in the figure: one population, characterized by M$_\mathrm{peak}\lesssim10^{9.3}$~M$_\sun$ (corresponding to UFD galaxies), are quenched uniformly early, regardless of infall time. More massive galaxies' quenching times are correlated with infall time. \textit{Bottom}: The HI mass at infall (to 2~R$_\mathrm{vir}$) for the same galaxies as in the top panel. Galaxies that are star-forming at infall are shown with squares, while galaxies that are quenched at infall are shown as triangles. Galaxies with less than $10^3$~M$_\sun$ in HI at infall are shown at the bottom of the figure. Combined with the top panel, the figure shows three populations of galaxies: galaxies that have lost their gas and quenched prior to infall, galaxies that have quenched but retained their gas prior to infall, and galaxies that are quenched after infall.}
    \label{fig:t90}
\end{figure}

To isolate the role of environment in UFD quenching, Figure~\ref{fig:t90} focuses on two processes pertaining to dwarf galaxies: star formation quenching and gas loss. The top panel shows the quenching time (here defined as $\tau_{90}$, the time when a galaxy reached 90\% of its final stellar mass) of all {\it quenched} galaxies as a function of infall time to 2~R$_\mathrm{vir}$\footnote{Previous work \citep[e.g.,][]{Behroozi2014, Fillingham2018} has found that environmental effects from the Milky Way extend out to ${\sim}2$~R$_\mathrm{vir}$, so we compare to infall at this radius. We have additionally confirmed that the results of Figure~\ref{fig:t90} hold true for infall radii between 1-3~R$_\mathrm{vir}$.}, colored by peak halo mass. If halos had multiple infalls, the time of their first infall is used. Halos which have never approached within 2~R$_\mathrm{vir}$ are assigned an infall time of 14~Gyr. We also mark the beginning and end of reionization as implemented in our simulations \mbox{($z=15$ - $6$)}.

Figure~\ref{fig:t90} shows two different galaxy populations---galaxies that quenched uniformly early regardless of infall time, and galaxies whose quenching correlates with infall. The populations are approximately separable by mass, and the division between these galaxies occurs at M$_\mathrm{peak}\sim10^{9.3}$~M$_\sun$\footnote{We have confirmed that the division in peak halo mass is the same in the Near Mint runs, and so quenching results in our simulations are resolution-independent.}. This division coincides with that of UFD galaxies, which have M$_\mathrm{peak}\lesssim10^{9.5}$~M$_\sun$. Though we do not show show it here, we have verified that the two populations of Figure~\ref{fig:t90} are just as clearly separated for infalls to 1~R$_\mathrm{vir}$ as for 2~R$_\mathrm{vir}$. Additionally, for small halos hosting UFD galaxies, quenching was generally earlier than infall to 3~R$_\mathrm{vir}$, let alone 1~R$_\mathrm{vir}$. Combined with the general lack of connection between infall time and quenching time, the early cessation of star formation indicates that reionization and/or supernova feedback was responsible for quenching the majority of the UFDs.
Larger halos, whose quenching is tied to infall, stop forming stars as a result of environmental effects. They are studied in more detail in \citet{Akins2020}.

Interestingly, the processes responsible for quenching are not necessarily the same processes that remove gas from the galaxy. The bottom panel of Figure~\ref{fig:t90} shows the HI mass at infall (to 2~R$_\mathrm{vir}$) for the same galaxies as in the top panel; any galaxy without gas or with HI mass $<10^3$~M$_\sun$ is shown at the bottom of the panel. Galaxies that are star-forming at infall are shown as squares, while galaxies quenched at infall are shown as triangles.

The figure shows that a large number of galaxies that are already quenched at infall have retained substantial amounts of cold gas (we have confirmed the same holds true for infalls to 1~R$_\mathrm{vir}$). The division for these galaxies occurs at M$_\mathrm{peak}\sim10^9$~M$_\sun$. In other words, while the large majority of UFDs quench early, well before any interaction with the Milky Way, UFDs residing in more massive halos ($10^{9.0}\leq\mathrm{M}_\mathrm{peak}\leq10^{9.5}$~M$_\sun$) retain their gas until they interact with the Milky Way. The processes responsible for quenching (supernova feedback and reionization) do not fully heat or remove cold gas prior to infall. Yet, in the present day, as we discuss in Section~\ref{sec:gasrichufds}, most of the UFDs that contained gas at infall no longer do.

In summary, we see a transition in the way reionization acts on halos as we increase our mass scale. UFDs in halos with M$_\mathrm{peak}\lesssim10^{9.0}$~M$_\sun$ are quenched uniformly early; the vast majority of these halos also lose their gas quickly. Galaxies in halos with \mbox{$10^{9.0}\leq\mathrm{M}_\mathrm{peak}\leq10^{9.5}$~M$_\sun$} represent a transition range, in which the galaxy is quenched early but can retain some halo gas for many Gyr, until infall. Above \mbox{M$_\mathrm{peak}\geq10^{9.5}$~M$_\sun$}, galaxies are quenched environmentally, if at all; these galaxies also retain gas until the present day, as we discuss below.

\section{Case Studies}\label{sec:casestudies}

Here we discuss several interesting dwarf galaxies; these either have unique properties or have interesting evolutionary histories. Taken together, they demonstrate how interactions in a Milky Way environment contribute to the diversity observed in faint dwarf galaxy properties.

\subsection{Gas-rich UFDs}\label{sec:gasrichufds}
\begin{figure}
    \epsscale{1.15}
    \plotone{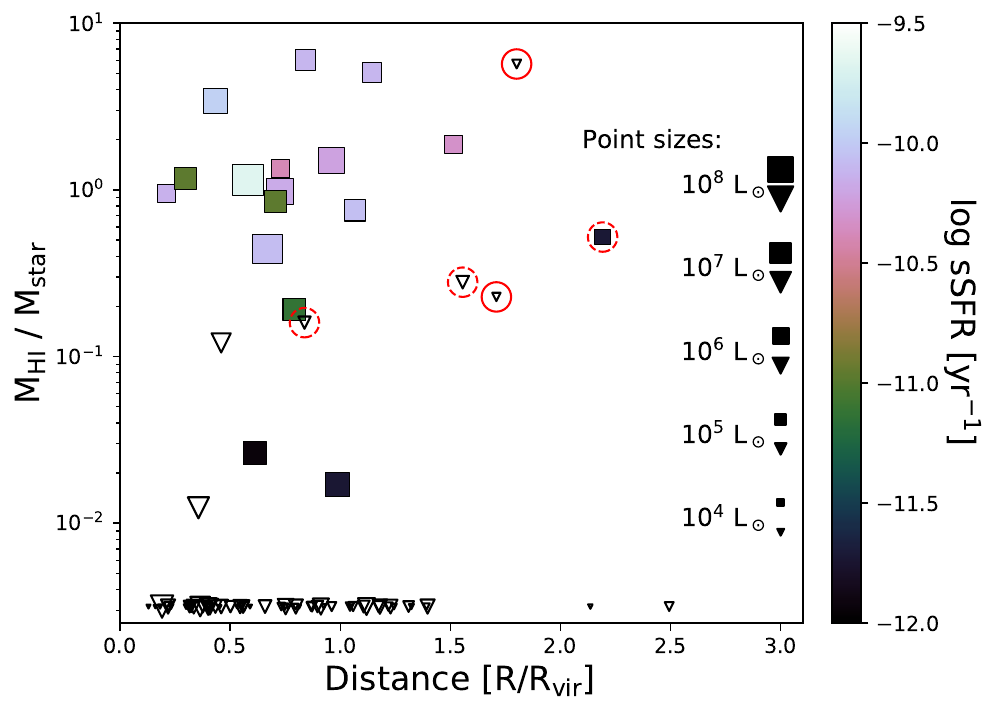}
    \caption{HI fraction as a function of galactocentric distance for all galaxies in the sample. Squares represent star forming galaxies, colored by specific star formation rate, while triangles represent quenched galaxies. Galaxies devoid of HI are placed at the bottom of the plot. Finally, squares and triangles are sized according to the galaxy's $V$-band luminosity, with references to guide the eye on the right-hand side of the plot. While most faint galaxies are quenched and have no gas or HI, a few have retained their gas. The solid circles highlight two UFDs that have non-zero HI masses, located at ${\sim}2$~R$_\mathrm{vir}$, while the dashed circles show slightly brighter galaxies ($-10< M_V<-8$) with HI.}
    \label{fig:hi}
\end{figure}

Figure~\ref{fig:hi} shows the HI fractions of all galaxies in the sample as a function of their distance from the Milky Way. Triangles at the bottom indicate galaxies devoid of HI. Points are sized by galaxy $V$-band luminosity; on the right are several points for comparison. Finally, the points are colored by specific star formation rate (sSFR; defined as SFR/M$_\mathrm{star}$), with quenched galaxies as unfilled points. Star formation rates are calculated as in \citet{Tremmel2019a}: we calculate the 25~Myr SFR, except in cases where two or fewer star particles form. To minimize numerical noise in these cases we use the average SFR over 250~Myr.

Most of the brighter galaxies are actively star forming, and even the few that are quenched (or nearly so) retain some HI. While the lowest sSFR galaxies are all found within 1 R$_{\rm vir}$, the higher sSFR galaxies can be found across a range of galactocentric distances. It is possible the brighter simulated galaxies are more HI-rich than their observed counterparts; among satellites, only the Magellanic Clouds around the Milky Way and NGC 185, NGC 205, IC 10, and LGS 3 around M31 are known to host HI \citep{McConnachie2012}. If the simulated brighter galaxies in Figure~\ref{fig:hi} are too HI-rich, it could be indicative of insufficient ram pressure stripping as they fall in and orbit the main halo. However, given that \citet{Akins2020} found the (ram pressure-induced) quenching timescales of satellites in the Near Mint simulations are consistent with observations, we find this unlikely. Additionally, results from the SAGA survey \citep{Geha2017, Mao2020} suggest the Local Group may have higher quenched fractions than typical Milky Way-like galaxies, which would indicate the Local Group satellites are less HI-rich than in a typical Milky Way-mass galaxy. Nonetheless, \citet{Akins2020} found that the different Milky Way-mass galaxies in the Near Mint simulations produce a variety of quenched fractions similar to those observed.

As seen before in Figure~\ref{fig:csfh}, all UFDs are quenched. However, not all the UFDs are devoid of HI. Two UFDs near 2~R$_\mathrm{vir}$ (about 600~kpc) have non-zero HI masses; they are shown in solid circles in Figure~\ref{fig:hi}. While unusual among UFDs, they are not as rare when considering slightly more massive galaxies; among galaxies with $-10<M_V<-8$, there are three galaxies with HI, shown in dashed circles. Most of them are concentrated near 2~R$_\mathrm{vir}$, but one of them is located within the virial radius. The more HI-rich UFD, with an HI mass of $3.5\times10^5$~M$_\sun$, should be detectable by surveys such as ALFALFA \citep{Giovanelli2010} or through targeted observations. The more massive HI-rich dwarf galaxies may likewise be detectable through targeted searches, including the nearest one, which has a distance of 200~kpc and HI mass of $4\times10^4$~M$_\sun$.

The standard view of UFD galaxies is not only that they quench during or shortly after the epoch of reionization, but that they are devoid of gas. Previous searches for HI in UFDs have yielded upper limits with no detections \citep[e.g.,][]{Grcevich2009, Spekkens2014, Westmeier2015, Crnojevic2016b}. Leo P ($M_V=-9.27$) is among the fainter known galaxies hosting HI, but at a distance of 1.62~Mpc \citep{Giovanelli2013, Rhode2013, McQuinn2015} it is far more isolated than most known UFDs. Currently, Leo T (whose luminosity of $M_V=-8.0$ is on the edge of the UFD definition; see, e.g., \citealt{Simon2019}) is the faintest known galaxy hosting HI \citep{Irwin2007, Ryan-Weber2008}; at 420~kpc from the Milky Way, it is also among the more distant of the known galaxies at such low luminosity.

\begin{figure}
    \epsscale{1.15}
    \plotone{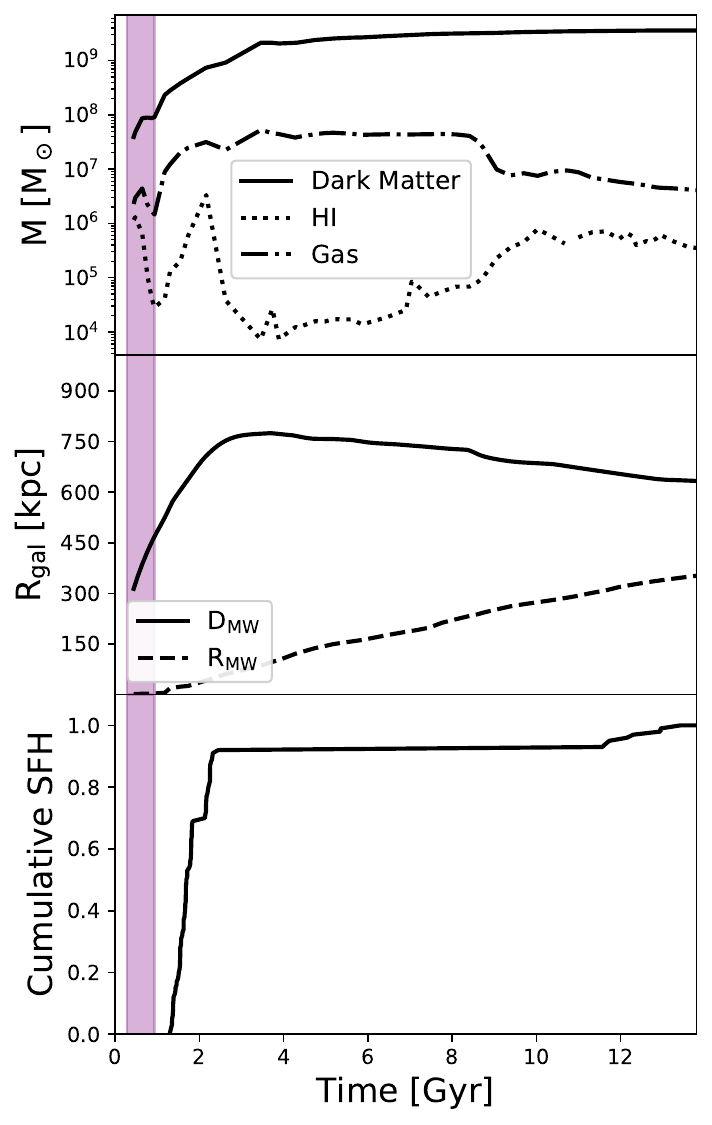}
    \caption{Evolutionary history of an HI-rich UFD. The top panel shows dark matter, total gas, and HI mass of the main progenitor through time. The middle panel shows its galactocentric distance (interpolated between time steps with a cubic spline) as well as the virial radius of the Milky Way-like galaxy Sandra. The bottom panel shows the cumulative star formation history. The purple band shows $z=15-6$, the epoch of reionization. Despite quenching early, the galaxy retained all of its gas until falling into ${\sim}2$~R$_\mathrm{vir}$. At this point, it began losing gas but continued gaining HI, as well as restarting star formation. Nonetheless, it retains appreciable HI content at $z=0$.}
    \label{fig:hicasestudy}
\end{figure}

Figure~\ref{fig:hicasestudy} shows the evolutionary history of the most HI-rich UFD galaxy in Figure~\ref{fig:hi} (halo 24 in Sandra). The top panel shows the dark matter, total gas, and HI mass of the main progenitor. The middle panel shows the galactocentric distance of the halo as well as the virial radius of the Milky Way-like main halo. For visualization purposes, the galactocentric distance is interpolated between time steps on a cubic spline, though as shown in \citet{Richings2020} this can lead to underestimates in pericentric distances. The bottom panel shows the cumulative SFH. The purple band indicates the duration of the epoch of reionization.

Figure~\ref{fig:hicasestudy} shows that the HI-rich UFD had a largely uneventful history; star formation began shortly after the epoch of reionization, and it formed the bulk of its stars around 1.5-2.5~Gyr, at which time it also lost most of its cold gas. It continued accreting dark matter throughout its lifetime, showing no obvious signs of interaction with either the Milky Way or other dwarf galaxies. However, during its long approach to the Milky Way, it began accumulating HI again, while also losing gas overall. When it neared 2~R$_\mathrm{vir}$, it restarted star formation, similarly to the reignited galaxies of \citet{Wright2019}. The increase in HI mass, decrease in total gas mass, and renewed star formation can all be explained by ram pressure on the infalling galaxy as it approaches the halo of the Milky Way-like host; ram pressure compresses the galaxy's gas, increasing gas densities and promoting star formation \citep[e.g.,][]{Fujita1999, Bekki2003, Du2019}.  We classify it now as quenched---it last formed stars 400~Myr ago---but it may be more accurately described as forming stars at a rate below our resolution, as the formation of star particles at such low SFR is subject to shot noise.

Recently, \citet{Janesh2019} found 5 candidate UFD galaxies in imaging follow-up to ultra-compact high-velocity clouds (UCHVCs) discovered in the ALFALFA HI survey \citep{Giovanelli2005}. Of the candidates, several have distances of ${\sim}2-3$ R$_\mathrm{vir}$, HI masses of ${\sim}10^{5-6}$~M$_\sun$, and estimated magnitudes of -4 to -7. Given their similar properties, the very faint yet gas-rich galaxies of Figure~\ref{fig:hi} may serve as simulated counterparts to these recently discovered candidate galaxies, and can offer insight into their origin. If these UFDs are analogs to observed UCHVCs, they would provide evidence that UCHVCs reside in their own dark matter halos \citep{Faerman2013}. We reserve for future work more detailed comparisons between the HI properties of the simulated UFDs and theoretical expectations.

\subsection{Late-forming UFD}\label{sec:lateformer}

Another outlier among the UFDs is the late-forming UFD of Figure~\ref{fig:csfh} (halo 409 in Elena). It not only began star formation later than 2~Gyr, it then continued forming stars for over 7~Gyr, albeit with some periods of quenching during that time. Both its late onset and long duration make it unusual among UFDs.

\begin{figure}
    \epsscale{1.15}
    \plotone{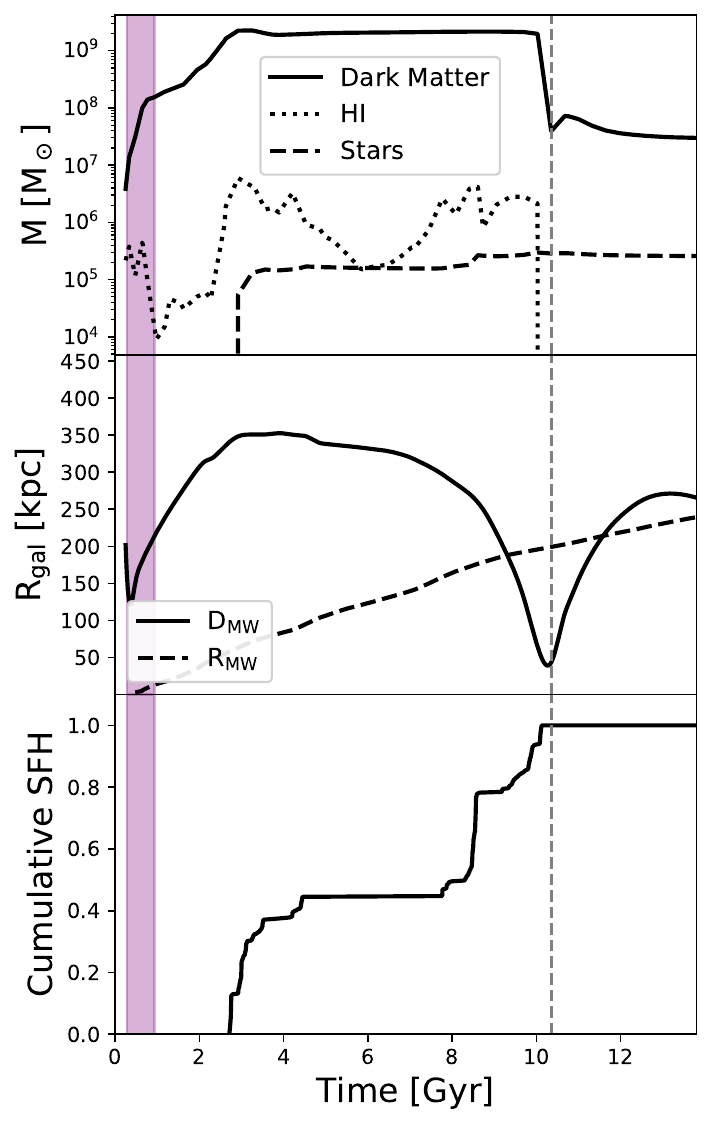}
    \caption{Evolutionary history of a late-forming UFD. The top panel shows dark matter, HI, and stellar mass of the main progenitor through time. The middle panel shows its galactocentric distance (interpolated between time steps with a cubic spline) as well as the virial radius of the Milky Way-mass host. The bottom panel shows the archaeological cumulative star formation history---i.e. the star formation history as inferred from the unstripped stellar population remaining at $z=0$. The purple band shows $z=15-6$, the epoch of reionization. The galaxy began forming stars late, and continued forming stars (with long pauses) until losing all of its gas and most of its dark matter during its first pericentric passage.}
    \label{fig:lateufdcasestudy}
\end{figure}

Figure~\ref{fig:lateufdcasestudy} shows this galaxy's evolutionary history, akin to Figure~\ref{fig:hicasestudy}. We additionally mark with a dashed vertical line the (approximate) time of pericenter during the halo's orbit. Interestingly, this halo began forming stars near apocenter, and continued forming stars (though with long pauses) until pericenter, at which point it quickly lost all of its gas and over 90\% of its dark matter. Similarly dramatic tidal stripping occurring near pericenter is common \citep[e.g.,][]{Klimentowski2009, Penarrubia2010}, particularly on highly eccentric orbits with close approaches, and in gas-rich dwarf galaxies where ram pressure stripping lowers the central density of the halo \citep[e.g.,][]{Kazantzidis2017}. We note that this galaxy is a backsplash galaxy; due to its eccentric orbit, it is currently in the near-field despite having a past pericentric passage closer than 50~kpc.

While Section~\ref{sec:quenching} showed that the bulk of the UFDs are quenched early on by reionization and/or feedback, this galaxy demonstrates that even these seemingly simple systems can exhibit a variety of histories. While most of the observed UFD star formation histories show early quenching (see Figure~\ref{fig:csfh}), galaxies such as this one may explain the later star formation observed in a couple of the \citet{Weisz2014a} UFDs. This kind of UFD is quite rare in our sample, however, so additional CMD-derived star formation histories are needed to better constrain how rare such dwarfs are in the observed Universe.

In addition to its unique star formation history, this galaxy's kinematics and structure are worth noting. With a half-light radius of 600~pc, $M_V=-7.9$, and a line-of-sight velocity dispersion of 3.2~km/s, this galaxy is in many ways an analog of Crater~2 \citep[$M_V=-8.2$;][]{Torrealba2016}, which is unusually large ($r_h\sim1$~kpc) and cold ($\sigma_v=2.7$~km/s). For halo 409, the same severe tidal stripping that quenched its star formation is also likely responsible for its low velocity dispersion; before tidal stripping, it had a velocity dispersion of 6.5~km/s, which while low, would not be rare. In Figure~\ref{fig:sigma_massloss}, this galaxy is one of the most severely tidally stripped, and also has one of the lowest velocity dispersions of any galaxy in the simulations.

Prior work using the APOSTLE simulations addressed the formation of cold, large galaxies such as Crater 2 \citep{Torrealba2018}, and similarly predicted that severe mass loss could explain their structure. However, as they could not directly probe such faint galaxies, they instead tied their derived stellar mass-halo mass relation with the tidal stripping evolutionary tracks of \citet{Errani2015} to infer progenitor properties from present-day dwarf galaxies. While our halo 409 exhibits a similar total mass loss to their predictions (${\sim}99$\%), they also predict similar tidal stripping in the stellar component. Halo 409, on the other hand, lost less than 10\% of its stars. This preferential stripping of the dark matter component may explain a lack of (so-far) observed tidal debris in the vicinity of Crater 2.

There are two additional Crater 2 analogs in the simulations, which similarly underwent severe tidal stripping: halo 1467 in Sandra has $M_V=-8.3$, $r_h=1.3$~kpc, and $\sigma_v=2.6$~km/s, and halo 2026 in Sandra has $M_V=-7.8$, $r_h=1.05$~kpc, and $\sigma_v=2.5$~km/s. These are among the most diffuse galaxies in our sample, with central surface brightnesses of ${\sim}30$~mag~arcsec${}^{-2}$. Unlike the above-discussed halo 409 in Elena, however, they have been stripped of the majority of their stars, and so would potentially have observable tidal debris in their vicinity.

\subsection{Compact Dwarf}\label{sec:ucd}

Below, we discuss a compact galaxy that forms in these simulations. Cosmological simulations, while successful in reproducing a wide array of dwarf galaxies, have had trouble simulating compact dwarf galaxies \citep[e.g.,][]{Jeon2017, Fitts2017, Revaz2018, Garrison-Kimmel2019}. In particular, none of these previous cosmological simulations have reproduced ultra-compact dwarf galaxies\footnote{\citet{Shen2014} did find, however, field galaxies with $M_\mathrm{star}\sim10^5$~M$_\sun$ and $r_h\sim80-90$~pc, which formed the bulk of their stars close to $z=0$, leading to their compact morphology.} \citep[UCDs; ][]{Hilker1999, Drinkwater2000, Phillipps2001}, a population of galaxies with M$_\mathrm{star}\sim10^6-10^8$~M$_\sun$ and $r_h\sim10-100$~pc, nor have they produced compact elliptical (cE) galaxies (M$_\mathrm{star}\sim10^8-10^{10}$~M$_\sun$ and $r_h\sim100-700$~pc), of which M32 is the prototype.

\begin{figure}
    \epsscale{1.15}
    \plotone{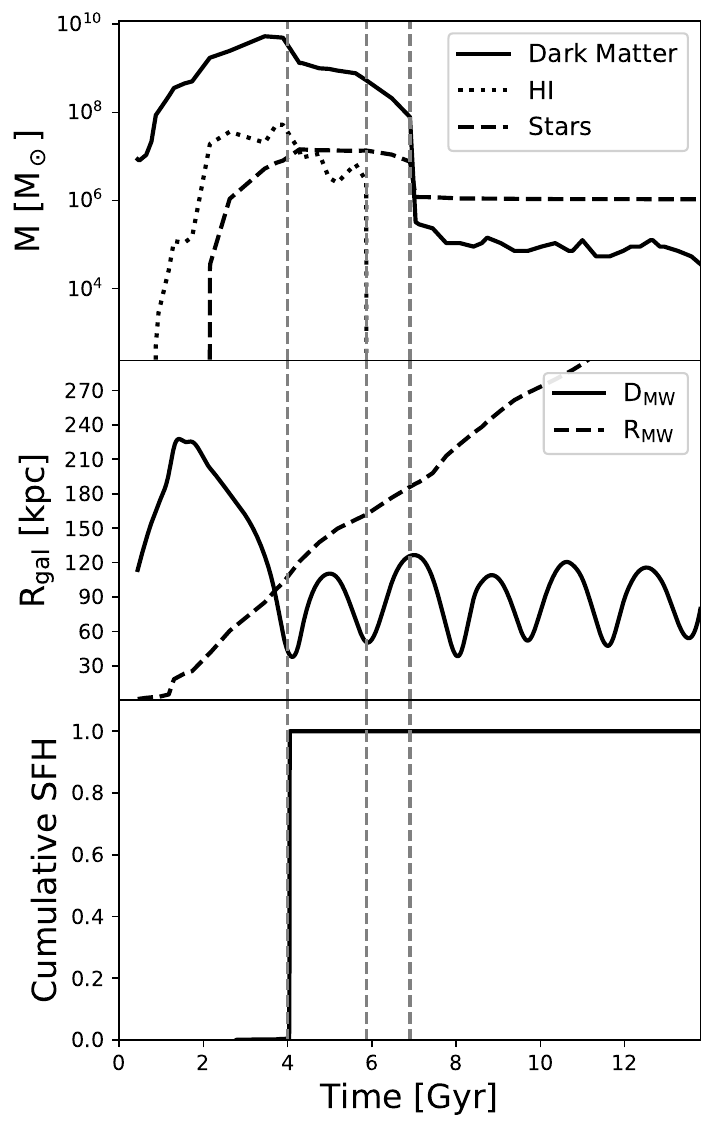}
    \caption{Evolutionary history of a compact dwarf galaxy. The top panel shows dark matter, HI, and stellar mass of the main progenitor through time. The middle panel shows its galactocentric distance (interpolated between time steps with a cubic spline) as well as the virial radius of the Milky Way-like host. The bottom panel shows the archaeological cumulative star formation history---i.e. the star formation history as inferred from the unstripped stellar population remaining at $z=0$. We mark using vertical dashed lines the time of greatest gas and dark matter loss (defined by steepest logarithmic slope), which correspond to the second pericenter and apocenter after infall, respectively. The archaeological star formation history differs from the true stellar mass through time because most of the stellar material has been stripped by the present day.}
    \label{fig:compacttrace}
\end{figure}

In Figure~\ref{fig:sizes}, there is a clear outlier in the size-luminosity plane. While hosting a typical $V$-band magnitude of $-9.2$, it is unusually compact, with a half-light radius of 40~pc. This size and luminosity places it firmly within the faint end of the UCD population \citep{Brodie2011}. As we discuss below, this galaxy (halo 1179 in Sandra) is the remnant of a severely tidally stripped dwarf galaxy.

Figure~\ref{fig:compacttrace} shows the galaxy's evolutionary history; the top panel shows the dark matter, gas, and stellar mass of the halo as a function of time, the middle panel shows the orbital history of the galaxy, and the bottom panel shows the archaeological cumulative star formation history---i.e. the star formation history as inferred from just the remnant stellar population. We also mark the times of most severe gas loss and dark matter loss. Interestingly, while the galaxy loses its gas at second pericenter, it loses its dark matter at the following apocenter. It may be that the halo is tidally shocked during its pericentric passage, resulting in heating and expansion that lead to greater susceptibility to tidal mass loss \citep[e.g.,][]{Gnedin1999a, Gnedin1999b}. Alternatively, ram pressure stripping at pericenter may have left the halo more susceptible to stripping \citep[e.g.,][]{Kazantzidis2017}. Ultimately, the tidal stripping was incredibly severe, with the galaxy losing all of its gas and over 99.99\% of its dark matter over the course of several Gyr\footnote{In the present day, this halo hosts $<10$ dark matter particles; nonetheless, these particles were sufficient to trace a main progenitor back in time. As an extra check, given the paucity of dark matter particles, we ensured that tracing the star particles separately yielded the same main progenitor halo.}. Currently, it would be observationally consistent with being devoid of dark matter. 

The bottom panel of Figure~\ref{fig:compacttrace} shows that the stars that constitute the central cluster formed within a very short period of time at ${\sim}4$~Gyr, at the time of the galaxy's first pericenter. These stars formed as a single, compact cluster, with approximately the same half-light radius as their present-day descendent. We find that the surviving stars all formed at the extreme high-pressure tail allowed by our star formation model \citep[see, e.g.,][]{Munshi2014}, which may explain their quick formation and initial compact nature. The summary of the formation scenario for this galaxy, then, is that it is the remnant of a large star cluster that formed in a typical, dark matter-dominated dwarf galaxy, that was then stripped of all dark matter, leaving only the compact, dense cluster as a dark matter-free galaxy.  Among the many UCD formation scenarios proposed, halo 1179's formation is most consistent with being the nuclear remnant of a tidally ``threshed'' dwarf galaxy \citep[e.g.,][]{Bassino1994, Bekki2001}.

However, this galaxy is at the edge of our resolution; in fact, with a gravitational softening length of 87~pc, the half-light radius is below our force resolution. It is therefore in some ways surprising that the galaxy is dynamically stable. This simulation allows for a minimum hydrodynamical smoothing length of 11~pc, making the gas clump that formed this cluster hydrodynamically resolved at the time of formation.  The structure of the stars has evolved little since formation, despite being below the force resolution. However, given its sub-resolution size, we are cautious that this galaxy may be influenced by unidentified numerical issues. Additionally, it is possible that the dark matter halo was stripped too efficiently due to artificial numerical disruption \citep{vandenBosch2018}.

Finally, we note that this galaxy was identified using AHF, which is tuned to find cosmological overdensities (see Section~\ref{sec:sims}), which is biased towards the prevailing dark matter-dominated galaxies. The only reason this galaxy was identified at all is due to its extremely high baryonic density. It is therefore likely that other dark matter-free and/or compact galaxies of slightly lower density are not being identified. Future work will return to these topics using alternate methods for identifying galaxies.

\section{Discussion}\label{sec:discussion}

We have presented a new set of cosmological hydrodynamic simulations of Milky Way-like galaxies, capable of resolving satellite and near-field galaxies down to $M_V\sim-4$. These simulations simultaneously produce realistic galaxies from the UFD regime to Milky Way mass, with the same feedback and star formation recipes for all galaxies.

\subsection{Comparison to Previous Works}\label{sec:previouswork}

\subsubsection{Structural Properties and Scaling Relations}

To date, several simulation groups have simulated dwarf galaxies in the same luminosity ranges as in this work, at comparable or higher resolution. However, none yet have done so in the environment of the Milky Way, which requires substantially more computational investment. Nonetheless, many of the prior simulations, like \citet{Simpson2013}, \citet{Onorbe2015}, \citet{Fitts2017}, \citet{Jeon2017}, \citet{Revaz2018}, and \citet{Agertz2020}, yield consistent results to many of our scaling relations, albeit each one doing so in a much narrower luminosity range. Much of this consistency likely results from the high dynamical mass-to-light ratios of these low mass dwarfs, which lead to the dark matter halos setting the structural properties of the galaxy \citep[e.g.,][]{Agertz2020}.

While our galaxy sizes and metallicities are consistent with those of other groups, our results are in some tension with those of \citet{Wheeler2019}. Their faintest galaxies are more diffuse (Figure~\ref{fig:sizes}) and less chemically enriched (Figure~\ref{fig:mzr}) than those in our simulations. There are several possible explanations for these discrepancies. Recently, \citet{Agertz2020} demonstrated that the mass-metallicity relation is highly sensitive to feedback strength. Explosive feedback can shut down star formation quickly and expel enriched gas, leaving stellar metallicities well below the observed relation. In their tests, the strongest feedback resulted in essentially primordial abundances. The feedback implementation in the FIRE-2 simulations is quite different from those implemented in \textsc{ChaNGa}.  While we have included only thermal energy from supernovae as feedback, the FIRE-2 feedback model \citep{FIRE-2} incorporates more feedback channels, including both energy and momentum injection from supernovae, radiation heating, and radiation pressure. Recently, \citet{Iyer2020} showed that dwarf galaxy SFHs are burstier in FIRE-2 than in \textsc{ChaNGa}, which may be a reflection of the different feedback implementations.

Most of the UFDs in \citet{Wheeler2019} are lower surface brightness than we find for our UFDs, but they are also at fainter luminosities than we are able to explore.  Thus, it is not clear if there is a discrepancy between our size results, but we note that all of the \citet{Wheeler2019} UFDs at these fainter magnitudes are larger than have been observed, despite having a gravitational force softening of only 14 pc.  Thus, we speculate on how the different feedback strengths might impact sizes. At higher masses, feedback has been shown to heat the stellar component \citep{Agertz2016, El-Badry2016, Chan2018}. Perhaps explosive outflows could also lead to large sizes in UFDs. However, the FIRE-2 simulations produce realistic sizes in higher mass dwarfs \citep[at lower resolutions;][]{Onorbe2015, Fitts2017}, indicating that if the feedback implementation is affecting galaxy sizes, it would be a resolution-dependent phenomenon.  Alternatively, \citet{Revaz2016} found that 2-body relaxation in their simulated galaxies led to a lack of compact dwarfs. \citet{Ludlow2020} recently demonstrated that gravitational softening lengths that are too small can exacerbate this issue and lead to greater galaxy sizes than with larger softening lengths.

Finally, we note that none of the FIRE-2 galaxies in \citet{Garrison-Kimmel2019} had velocity dispersions below 5~km/s, despite capturing the same environmental processes that in our simulations lead to dispersions as low as ${\sim}2$~km/s. It is possible that dynamical heating from 2-body interactions plays a role. Alternatively, stars may be born too kinematically hot, rendering it difficult to lower the velocity dispersion below 5~km/s even with tidal stripping. In fact, \citet{Sanderson2020} recently showed that, despite forming in dense, self-shielding gas, the youngest stars in the \mbox{FIRE-2} Milky Way simulations have (total) velocity dispersions $\gtrsim20$~km/s higher than those observed in the Milky Way (see their Figure 2).  Likewise, \citet{El-Badry2016} and \citet{Yu2020} showed that some stars in FIRE-2 are born in feedback-driven superbubbles with large initial radial velocities.

\subsubsection{Quenching}

Our results indicating that most UFDs were likely quenched by reionization (and feedback) are in line with previous cosmological simulations of field dwarf galaxies \citep[e.g.,][]{Simpson2013, Munshi2013, Munshi2017, Munshi2019, Wheeler2015, Wheeler2019, Jeon2017,  Revaz2018, Rey2019, Rey2020}. We note that \citet{Rey2020} also find that above \mbox{M$_\mathrm{peak}\sim10^9$~M$_\sun$} dwarf galaxies quenched by reionization can remain gas-rich. While in our simulations these galaxies generally lose their gas later as they fall in to the Milky Way, \citet{Rey2020} find that in the field these galaxies can continue to accrete gas, and even reignite their star formation. This presents a possible second mechanism for restarting star formation, in addition to the ram pressure-induced star formation discussed in Section~\ref{sec:gasrichufds} or \citet{Wright2019}.

In this work, we have not separated out the contributions of feedback and reionization in UFD quenching at high redshift. Many prior works have relied on the timing, as we have here, to infer that reionization is primarily responsible for quenching. To separate the effects of the two processes, \citet{Jeon2017} instead resimulated a field UFD with energetic supernova feedback turned off. They found that the UFD did not quench in the latter run, implying that while reionization is important in quenching, feedback is also a necessary contributor.

On the other hand, prior works using fully coupled radiation-hydrodynamics simulations of the early Universe have found that reionization quenches galaxies residing in small halos, and that these galaxies do not quench in the same simulations run without radiative transfer or a UV background \citep{Ocvirk2016, Ocvirk2020, Katz2020}. Further, \citet{Katz2020} found that even in halos that do not form stars at all, outside-in reionization causes a net outflow of gas, which does not occur in their simulation without reionization. At higher masses, however, supernova feedback begins to contribute to the outflow rate from halos. Unfortunately, it appears that the importance of reionization versus supernova feedback may be dependent on the specific feedback implementation, so we cannot assume results from their simulations would hold true in ours. We leave separating the effects of the two processes to future work.

Observationally, reionization quenching is consistent with previous works that compared infall times from dark matter-only simulations with CMD-derived star formation histories \citep[e.g][]{Rocha2012, Weisz2015, Rodriguez_Wimberly2019, Fillingham2019}. Recent work deriving UFD orbits using \textit{Gaia} proper motions also shows that many UFDs likely had later infalls than quenching times \citep[e.g.,][]{Fritz2018, Simon2018}. Recently, \citet{Miyoshi2020} directly compared the integrated orbital histories of several UFDs with the peaks in their inferred star formation histories. Unlike earlier works that use a static potential, they explicitly modeled the growing mass and radius of the Milky Way (compare, e.g., Figure~\ref{fig:lateufdcasestudy} to their Figure~1). They also found that star formation in UFDs occurs well before infall. However, they did find evidence that one UFD (CVn~I) had a second burst of star formation at infall, suggesting that some ultra-faint dwarf galaxies retained gas until infall, as we found in this work.  This example emphasizes the need for more observed UFD star formation histories in order to quantify how much variety there is, if any, in UFD star formation.

\subsection{Caveats}

While these simulations represent a step forward in the modeling of galaxies---particularly dwarf and ultra-faint dwarf galaxies---there remain limitations that this work, as in all simulations, still face.

\subsubsection{Reionization Model}

In these simulations, we adopted the uniform background UV photoionization and photoheating rate of \citet{Haardt2012}. This model has been shown to spuriously heat the IGM too early \citep{Onorbe2017}. Since we focus in this work on faint galaxies that are often quenched during reionization, correcting for a later reionization model may have a particularly large effect on the SFHs presented in  Section~\ref{sec:quenching}. More recent UV background models \citep[e.g.,][]{Puchwein2019} have corrected for this discrepancy, but the simulations presented in this work were begun before their release. On the other hand, because most UFDs infall to their parent halo much later than $z=6$, we do not expect a later reionization model to alter our conclusions about the source of quenching. Additionally, the overdense Milky Way environment may be better represented by an earlier reionization \citep{Li2014}. Indeed, \citet{Ocvirk2020} found that reionization suppresses star formation earlier in overdense regions than underdense ones.

\citet{Garrison-Kimmel2019b} found that using a later reionization model in their simulations shifts the majority of the star formation to even earlier times, as more stars are allowed to form in the pre-reionization era; if the same were true in our simulations, it would not change any of our main results. However, the specific mass at which galaxies transition from reionization and feedback quenching to environmental quenching may change, and galaxies in the UFD range could potentially form more stars before reionization ends, shifting them to higher masses. Future work will explore the impact of changing the reionization model in \textsc{ChaNGa}.

With the exception of a model for tracking Lyman-Werner radiation \citep{Christensen2012}, there is also no radiative transfer (RT) in these simulations. Ideally, reionization would be simulated self-consistently with RT rather than imposed as a uniform background. Radiative transfer, however, is computationally expensive, and simulations relying exclusively on RT without a cosmic UV background have been largely stopped at high redshift \citep[e.g.,][]{Wise2014, Gnedin2014, O'Shea2015, Pawlik2017, Rosdahl2018}, both due to the expenses of RT and the need to resolve cosmologically representative volumes.

\subsubsection{Resolution}\label{sec:resolution}

\begin{figure}
    \epsscale{1.2}
    \plotone{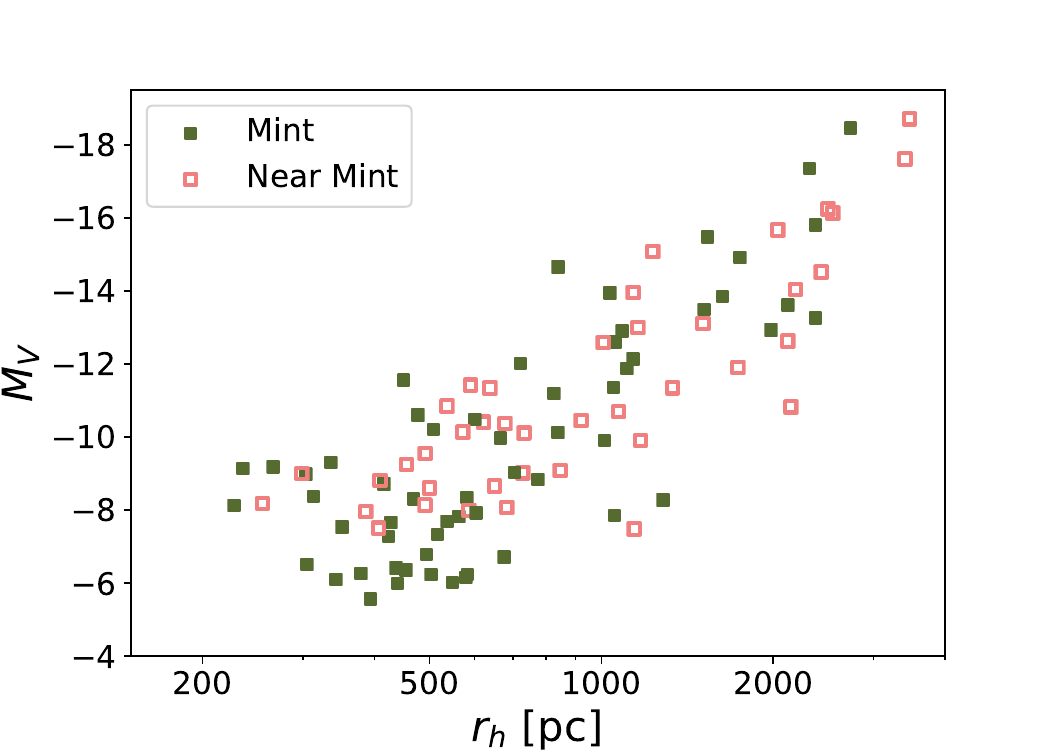}
    \caption{Size-luminosity relationship of galaxies in the Mint runs (presented in this work) and the Near Mint runs, at 2x lower spatial resolution. We exclude from this plot the UCD discussed in Section~\ref{sec:ucd}. The two resolutions largely span the same space, though fainter than $M_V\sim-8$ there may be indications of heating from 2-body interactions in the Mint runs.}
    \label{fig:resolution}
\end{figure}

A variety of recent works have shown that low-mass galaxies tend to approach a minimum size, which grows with time even for quiescent galaxies \citep[e.g.,][]{Furlong2017, Revaz2018, Pillepich2019}. A likely contributor to this behavior is numerical: the 2-body relaxation of unequal mass particles, such that the more massive (dark matter) particles sink to the bottom of the potential well and the less massive (star) particles slowly diffuse outward \citep[e.g.,][]{Binney2002, Ludlow2019a}. When approaching a simulation's resolution limits and studying poorly resolved galaxies, this can set a floor on the size of a galaxy and prevent, for example, the modeling of ultra-compact dwarf galaxies \citep[e.g.,][]{Garrison-Kimmel2019}.

We expect spurious dynamical heating of the stellar component to be more severe for larger mass ratios \citep{Ludlow2019a}. Since star particles in these simulations are smaller than the gas particles, which themselves are initially smaller than the dark matter particles by the ratio $\Omega_\mathrm{bar}/\Omega_\mathrm{DM}$, it is possible that 2-body interactions are even more impactful in these simulations than some others. However, as evidenced in Figure~\ref{fig:sizes} and others, we are indeed resolving effectively fairly compact galaxies.

Using collisionless simulations, \citet{Ludlow2019b} argue that halos are resolved above a convergence radius $r_\mathrm{conv}\approx0.055\times l$, where $l$ is the mean inter-particle spacing, and $l=L/N^{1/3}$, where $L$ is the simulation box size and $N$ is the number of particles. \citet{Ludlow2019a} find that spurious growth due to 2-body interactions is confined largely to radii below $r_\mathrm{conv}$. The dependence on softening length is weak, so long as the softening length is smaller than the convergence radius. Convergence, therefore, is based almost exclusively on particle number. The high resolution region starts with equivalent resolution to a $6144^3$ grid, which would yield a convergence radius of $r_\mathrm{conv}\approx450$~pc. If including baryons in the mean inter-particle spacing, the convergence radius reduces to $\lesssim300$~pc. Even so, Figure~\ref{fig:sizes} shows that several galaxies have half-light radii smaller than this value, though caution should be used in interpreting these particular galaxies.

To reveal any resolution effects, we show in Figure~\ref{fig:resolution} the size-luminosity relationship of the Mint and Near Mint runs, with half-light radii calculated using the 2D elliptical fits of Section~\ref{sec:structural}. For the most part, both resolutions span the same space. Surprisingly, the Near Mint runs, which have 175~pc resolution, are able to produce galaxies about as compact as the Mint runs, at 87~pc resolution. At $M_V\gtrsim-8$, below the resolution limit of the Near Mint runs, we see that Mint galaxies do not become more compact with decreasing luminosity. This may indicate heating from 2-body interactions. Indeed, the Near Mint runs have ``super-sampled'' dark matter, such that the mass ratio between dark matter and gas particles is about 3.5 times smaller than the Mint runs. It is possible that the Near Mint runs, therefore, suffer from less 2-body heating, and may explain why the Near Mint runs produce galaxies as small as the Mint runs.

\subsubsection{Feedback Models}

In this work, star particles were treated as simple stellar populations, in which the deposited supernova energy is calculated by integrating the IMF to calculate the number of exploding stars. As simulations increase in resolution, however, this methodology becomes insufficient, as such small stellar populations may contain only a few stars that explode as supernovae. It therefore becomes necessary to stochastically sample from the IMF. \citet{Applebaum2020} showed that while using a stochastic IMF has no effect on more massive galaxies \citep[see, however,][]{Su2018}, for galaxies in the UFD range stellar feedback becomes more effective and their stellar masses are reduced. However, there appears to be no change in the metallicity of the galaxies.

The simulations in this work were not run with a stochastic IMF, and so it is possible that with the more realistic feedback model the UFDs would be generally less massive. Additionally, it is in principle possible that the burstier feedback could lead to structural changes (e.g., by producing stronger repeated gas outflows). However, at the low masses of the UFDs, it is likely that galaxy morphology is set primarily by the dark matter halo properties and assembly history \citep{Rey2019, Agertz2020}. Additionally, no conclusions regarding the quenching of the ultra-faint dwarf galaxies would change, since any change in feedback from a stochastic IMF would lead toward even earlier quenching times.

Future work will additionally explore the impacts of different feedback models, including a superbubble model of supernova feedback that includes thermal conduction and models the subgrid multiphase ISM \citep{Keller2014}.  Preliminary work has shown that dwarf galaxy properties remain nearly identical between the blastwave and superbubble feedback implementations, but that stellar masses in Milky Way-mass galaxies are suppressed with superbubble feedback relative to blastwave by a factor of 2 to 3 \citep{Keller2015}.

\section{Summary}\label{sec:summary}

We have introduced a new suite of cosmological hydrodynamic zoom-in simulations, the DC Justice League simulations, run at ``Mint'' (87~pc) resolution with the \textsc{ChaNGa} N-Body + SPH code, and focusing on Milky Way-like environments. This suite has the highest-ever published mass resolutions for cosmological Milky Way-like simulations run to $z=0$, and pushes the boundaries of resolution forward to move beyond the classical dwarf regime, and begin the study of fainter dwarfs like those rapidly being discovered by digital surveys. With these simulations, we study a sample of 86 galaxies with $M_V\lesssim-3$, out to a distance of 2.5~R$_\mathrm{vir}$, including satellite and near-field galaxies.

We first compared our new galaxies to observations, ensuring that they are realistic and representative, and showed that our galaxy formation models continue to explain observations down into the UFD range. We found that the two simulations presented here, Sandra and Elena---whose galaxies span approximately 6 dex in luminosity, excluding the central Milky Way---reproduce the observations for a variety of scaling relations. In particular, with the exception of the compact ellipticals like M32, these galaxies span the full range of luminosities for a given size, down to ${\sim}200$~pc in half-light radius (Figure \ref{fig:sizes}). The galaxies also span the full range of observed kinematics of dispersion-supported systems (Figure \ref{fig:kin}), with tidal stripping responsible for the dynamically coldest galaxies. Given their central surface brightnesses, we predict all nearby galaxies will be observable by the Vera Rubin Observatory's co-added LSST.

We found our metallicities are generally consistent with observations for all galaxies with $L_V\gtrsim10^4$~L$_\sun$ (Figure \ref{fig:mzr}). The faintest galaxies are under-enriched in Fe compared to observed dwarfs, but the discrepancy disappears if total metallicity is considered.  This result suggests that the galaxies are forming and retaining metals but may be Fe poor due to either SFHs that are truncated before enrichement by SN\,Ia, or a lack of a model for ``prompt'' SN\,Ia in our simulations. Future work will investigate the discrepancy further.

We took advantage of the high resolution of the simulations to investigate the star formation and quenching of UFDs ($M_V\gtrsim-8$) in Section~\ref{sec:quenching}. We found that their SFHs are largely consistent with the limited number of available CMD-derived SFHs \citep{Brown2014, Weisz2014a}. Additionally, while the large majority of UFDs quench uniformly early and long before infall, many of the quenched UFDs still retain their gas until later interactions with the Milky Way.

In Section \ref{sec:casestudies}, we also highlighted dwarf galaxies that are the first of their kind to be simulated around a Milky Way-mass galaxy. One of them is an HI-rich UFD, which is atypical in our simulations, and unseen in observations of UFDs near the Milky Way. We find that while quenched early on, this galaxy retained its gas until its first infall towards the Milky Way, at which point it also briefly restarted forming stars. We also highlighted a late-forming UFD that is structurally similar to Crater 2, which both started forming stars late, and maintained ongoing star formation for many Gyr before quenching during a close pericentric passage. Finally, we also examined a compact, dark matter-free dwarf galaxy in our simulations, which formed as the remains of a tidally threshed galaxy. This galaxy may serve as an analog to the observed ultra-compact dwarf galaxies. These rare but unusual galaxies emphasize the need for additional observations (such as CMD-derived star formation histories or targeted HI observations) that can quantify the full diversity of faint dwarf galaxies.

The simulations we have introduced in this work demonstrate that a unified set of physics can simultaneously explain galaxy formation across many orders of magnitude, as well as naturally reproduce the variety seen in observations. The simulations show that one of the primary drivers of the variety seen in nearby galaxies is due to interaction with the Milky Way galaxy. These simulations fill a gap in the available literature, extending the study of dwarf galaxies around the Milky Way below the classical dwarf regime, which before was only accessible in field environments very different from the majority of the observations.

\acknowledgments
The authors would like to thank the anonymous referee for insightful comments that improved this work. We also thank Hollis Akins, Jillian Bellovary, Martin Rey, and John Wu for meaningful discussions relating to this work. EA and AMB acknowledge support from NSF grant AST-1813871. EA acknowledges support from the National Science Foundation (NSF) Blue Waters Graduate Fellowship. CRC acknowledges support from the NSF under CAREER grant No. AST-1848107. S. Shen acknowledges the support from the Research Council of Norway through NFR Young Research Talents Grant 276043. MT acknowledges the generous support of the Yale Center for Astronomy and Astrophysics Postdoctoral Fellowship. Resources supporting this work were provided by the NASA High-End Computing (HEC) Program through the NASA Advanced Supercomputing (NAS) Division at Ames Research Center. This research is part of the Blue Waters sustained-petascale computing project, which is supported by the National Science Foundation (awards OCI-0725070 and ACI-1238993) and the state of Illinois. Blue Waters is a joint effort of the University of Illinois at Urbana-Champaign and its National Center for Supercomputing Applications.  This research also made use of the NSF supported Frontera project operated by the Texas Advanced Computing Center (TACC) at The University of Texas at Austin.


\bibliography{biblio}{}
\bibliographystyle{aasjournal}



\end{document}